\documentclass{article}

\usepackage{jinstpub} 
\usepackage{inputenc,graphicx}
\usepackage{siunitx}
\usepackage{multirow}
\usepackage{makecell}

\newcommand{\monolith}{MONOLITH} 
 
\newcommand{\pd}{P_{\it density}} 
\newcommand{\toa}{\it TOA} 
 
\newcommand{\sigdtoa}{\sigma_{\Delta_{\it TOA}}} 
\newcommand{\sigdtoad}{\sigma_{\Delta_{\it TOA,d}}} 
\newcommand{\dtoa}{\Delta_{\it TOA}}
\newcommand{\dtoaid}{\Delta_{\it TOA,i, d}} 
\newcommand{\sigdut}{\sigma_{\it DUT}}
\newcommand{\sigmcpzero}{\sigma_{\it MCP0}}
\newcommand{\sigmcpone}{\sigma_{\it MCP1}}

\title{20 ps Time Resolution with a Fully-Efficient Monolithic Silicon Pixel Detector without Internal Gain Layer}

\author[a,1]{S. Zambito,\note{Corresponding author.}}
\author[a]{M. Milanesio,}
\author[a]{T. Moretti,}
\author[a,b]{L. Paolozzi,} 
\author[a]{M. Munker,}
\author[a]{R. Cardella,}
\author[a]{T. Kugathasan,}
\author[a]{F. Martinelli,}
\author[a,b]{A. Picardi,}
\author[c]{M. Elviretti,}
\author[c]{H. Rücker,}
\author[c]{A. Trusch,}
\author[a]{F. Cadoux,}
\author[a,2]{R. Cardarelli\note{Also at INFN Section of Roma Tor Vergata, Via della ricerca scientifica 1, Roma, Italy.},}
\author[a]{S. Débieux,}
\author[a]{Y. Favre,}
\author[a]{C. A. Fenoglio,}
\author[a]{D. Ferrere,}
\author[a]{S. Gonzalez-Sevilla,}
\author[a]{L. Iodice,}
\author[a,b]{R. Kotitsa,}
\author[a]{C. Magliocca,}
\author[a,b]{M. Nessi,}
\author[a]{A. Pizarro-Medina,}
\author[a]{J. Sabater Iglesias,}
\author[a]{J. Saidi,}
\author[a]{M. Vicente Barreto Pinto}
\author[a,1]{and G. Iacobucci}
\affiliation[a]{D\'epartement de Physique Nucl\'eaire et Corpusculaire (DPNC),
University of Geneva, 24 Quai Ernest-Ansermet, CH-1211 Geneva 4, Switzerland}

\affiliation[b]{CERN, CH-1211 Geneva 23, Switzerland}

\affiliation[c]{IHP — Leibniz-Institut für innovative Mikroelektronik, Im Technologiepark 25, Frankfurt (Oder), Germany}

\emailAdd{stefano.zambito@unige.ch, giuseppe.iacobucci@unige.ch}

\abstract{
A second monolithic silicon pixel prototype was produced for the MONOLITH project. The ASIC contains a matrix of hexagonal pixels with 100 µm pitch, readout by a low-noise and very fast SiGe HBT frontend electronics. Wafers with 50 µm thick epilayer of 350 $\Omega$cm resistivity 
were used to produce a fully depleted sensor. 
Laboratory and testbeam measurements of the analog channels present in the pixel matrix show that the sensor has a 130 V wide bias-voltage operation plateau at which the efficiency is 99.8\%.
Although this prototype does not include an internal gain layer, the  design optimised for timing of the sensor and the front-end electronics provides a time resolutions of 20 ps.
}

\begin{document}


\maketitle
\section{Introduction}
\label{sec:intro}

Monolithic Active Pixel Sensors (MAPS)~\cite{peric}, which contain the sensor in the same CMOS substrate utilised for the electronics, are particularly appealing since they offer all the advantages of industrial standard CMOS processing, avoiding the production complexity and high cost of the bump-bonded hybrid pixel sensors that are commonly used in particle-physics experiments.
Today MAPS represent a mature technology that matches the performance of hybrid silicon pixel sensors. Indeed MAPS are already used in a large LHC experiment~\cite{alice}.

The large  pile-up of events expected during the High-Luminosity LHC program at CERN requires timing capabilities of few tens of picosecond~\cite{Sadrozinski_2017}. 
This level of timing will be  achieved in the ATLAS~\cite{atlasTDR} and CMS~\cite{cmsTDR} upgraded detectors by timing layers with a coarse spatial granularity of  approximately 1 mm. 
In parallel with this established technology, the particle-physics community is trying to develop for future projects silicon sensors with high spatial resolution and, at the same time, the same level of timing capabilities. A recent review of the present efforts in this context can be found in~\cite{cartiglia}.

This research group is trying to develop MAPS with picosecond time capabilities. Making use of the commercial SG13G2 IHP 130 nm process \cite{SG13G2}, we produced a series of monolithic prototypes with very fast and low noise SiGe HBT front-end electronics that achieved time resolutions down to 36 ps~\cite{TTPET1,TTPET2,hexa_50ps,Paolozzi_2020,Iacobucci:2021ukp} with standard PN junction sensors without internal gain layer. 
This research line evolved
in the MONOLITH H2020 ERC Advanced project~\cite{monolith}, which exploits the novel multi-PN junction PicoAD sensor~\cite{PicoADpatent} to achieve picosecond-level time resolutions by means of the signal-over-noise boost provided by a continuous deep gain layer.
The results obtained with the PicoAD proof-of-concept monolithic prototype are reported
in~\cite{PicoAD_gain} and~\cite{PicoAD_TB}. 

Recently, the second monolithic silicon pixel matrix prototype for the MONOLITH project was  produced in the SG13G2 IHP process. The ASIC contains an evolution of the front-end electronics of~\cite{Iacobucci:2021ukp} and is designed for improved operation capabilities.
While the special PicoAD wafers that implement a gain layer were being manufactured, a version with a standard PN-junction sensor was produced using wafers with  50 µm thick epilayer of a resistivity of 350 $\Omega$cm.
In this paper we present the testbeam results obtained with this second MONOLITH prototype without  internal gain layer.
\section{ASIC Description and Characterisation}
\label{sec:asic}

\begin{figure}[!htb]
\centering
\includegraphics[width=.45\textwidth,angle=90,trim=0 0 0 0]{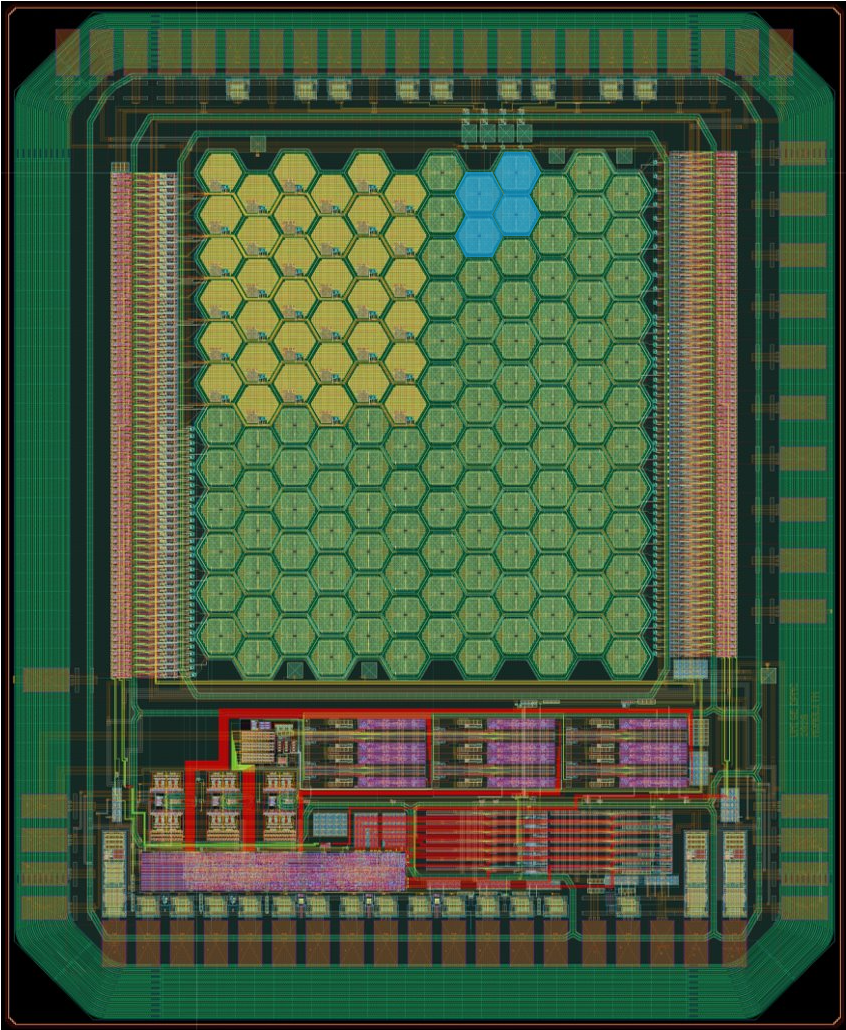}
\caption{\label{fig:ASIC} The monolithic pixel ASIC prototype. The pixel matrix consists of 144 hexagonal pixels of 65 µm side. Four analog pixels, highlighted in blue, have the preamplification stage directly connected to an analog driver with differential output.
The readout and configuration electronics is visible in the right side of the pixel matrix. The front end of the digital pixels are placed at the top and bottom sides of the matrix. The high voltage is applied at the back of the sensor and on the top surface outside the guard ring.
}
\end{figure}

The ASIC  prototype 
is a monolithic silicon pixel detector  in  SiGe BiCMOS technology, optimized for precise timing measurement. The ASIC was produced using the 130 nm SG13G2 process by IHP Microelectronics. It represents an evolution of a previous MONOLITH prototype~\cite{Iacobucci:2021ukp}. The ASIC matrix, shown in Figure~\ref{fig:ASIC}, contains 144 hexagonal pixels of 65 µm side (corresponding approximately to a pixel pitch of 100 µm), comprising 4 analog pixels (named hereafter OA0, OA1, OA2 and OA3) with the preamplifier directly connected to an analog driver to be read by an oscilloscope. 

In this second prototype the front-end electronics was improved to address the issues observed in the previous one. 
First of all, 
to remove the feedback path that was causing cross-talk between the channels and instability~\cite{Iacobucci:2021ukp},
the power supply of the preamplifier was  separated from that of the driver stage. This optimization had a positive impact on the charge measurement used for the time-walk correction and on the overall performance of the detector.  In addition, the driver stage was improved with the introduction of a differential analog output, which allows for a more stable operation of the amplifier at high gain of the frontend. 
The new driver allowed also for a reduction of the signal rise time, yielding a significant improvement of the detector timing jitter. 

To improve the sensor, low-resistivity  wafers with a 50 µm-thick  epitaxial layer of 350 $\Omega$cm resitivity were used. Laboratory measurements show that the epitaxial layer is fully depleted at a sensor bias voltage $HV$ = 70 V. 
An improved design of the  guard ring, combined with the increased resitivity of the epitaxial layer,   allows operation of the sensor up to $HV$ = 300 V.
This large bias-voltage operation plateau assures that the charge drift velocity can be saturated,  maximizing the timing response of the sensor.

A radioactive $^{55}$Fe source was used to characterise the ASIC analog electronics chain, comprising the pixel, the preamplifier and the driver,  at  five  working points corresponding to power-density values $\pd$ = 0.04, 0.13, 0.36, 0.9 and 2.7 W/cm$^2$.
The differential analog output signals 
were used for this measurement.
The equivalent noise charge (ENC) of the front-end electronics, reported in Table~\ref{tab:gain}, was computed as ENC = $\sqrt{\sigma_V^2 -2~\!\sigma_{\it scope}^2}/A_q$, 
where $\sigma_{V}$ is the standard deviation of the electronic-noise distribution measured by the oscilloscope when the front-end is connected,
 $\sigma_{\it scope}$ is  the voltage noise of the oscilloscope with the open input connector, taken twice to account for the signals of the two polarities,
 and $A_q$ is the charge gain.

\begin{table}[!htb]
\centering
\renewcommand{\arraystretch}{1.3}
\begin{tabular}{|c|c|c|c|c|}
\cline{1-5}
\cline{1-5}
\ $P_{\it density}$ [W/cm$^2$] &  $\sigma_V$ [mV] & $\sigma_{\it scope}$ [mV] & $A_q$ [mV/fC] & ENC [electrons] \\
\cline{1-5}
2.7   & ~~$ 1.45 \pm 0.02 $~~  & ~~$ 0.39\pm0.01 $~~ & ~~~$89.4 \pm 1.1 $~~~ & $~~93\pm2$ \\
0.9   & $ 1.06 \pm 0.02 $ & $ 0.39\pm0.01 $ & $67.4 \pm 0.8 $ & $~~84\pm1$ \\
0.36  & $ 0.89 \pm 0.01 $ & $ 0.39\pm0.01 $ & $45.3 \pm 0.8 $ & $~~97\pm2$ \\
0.13  & $ 0.58 \pm 0.01 $ & $ 0.16\pm0.01 $ & $26.6 \pm 0.5 $ & $ 125\pm5$ \\
0.04  & $ 0.67 \pm 0.01 $ & $ 0.16\pm0.01 $ & $27.0 \pm 0.5 $ & $ 145\pm8$ \\

\cline{1-5}
\end{tabular}
\caption{Charge gain and ENC measured with a $^{55}$Fe source at five different values of the front-end power density. The standard deviations  of the voltage noise and of the oscilloscope noise used for the calculation of the ENC are also reported.}
\label{tab:gain} 
\end{table}

\section{Testbeam Setup, Track-Selection Criteria and Data Samples}
\label{sec:setup}

A pion beam of $\SI{120}{\giga\electronvolt}$/c momentum delivered at the CERN SPS testbeam facility was used to characterise the MONOLITH 2022 ASIC prototype and measure its detection efficiency and time resolution. 

\begin{figure}[!htb]
\centering
\includegraphics[width=.98\textwidth]{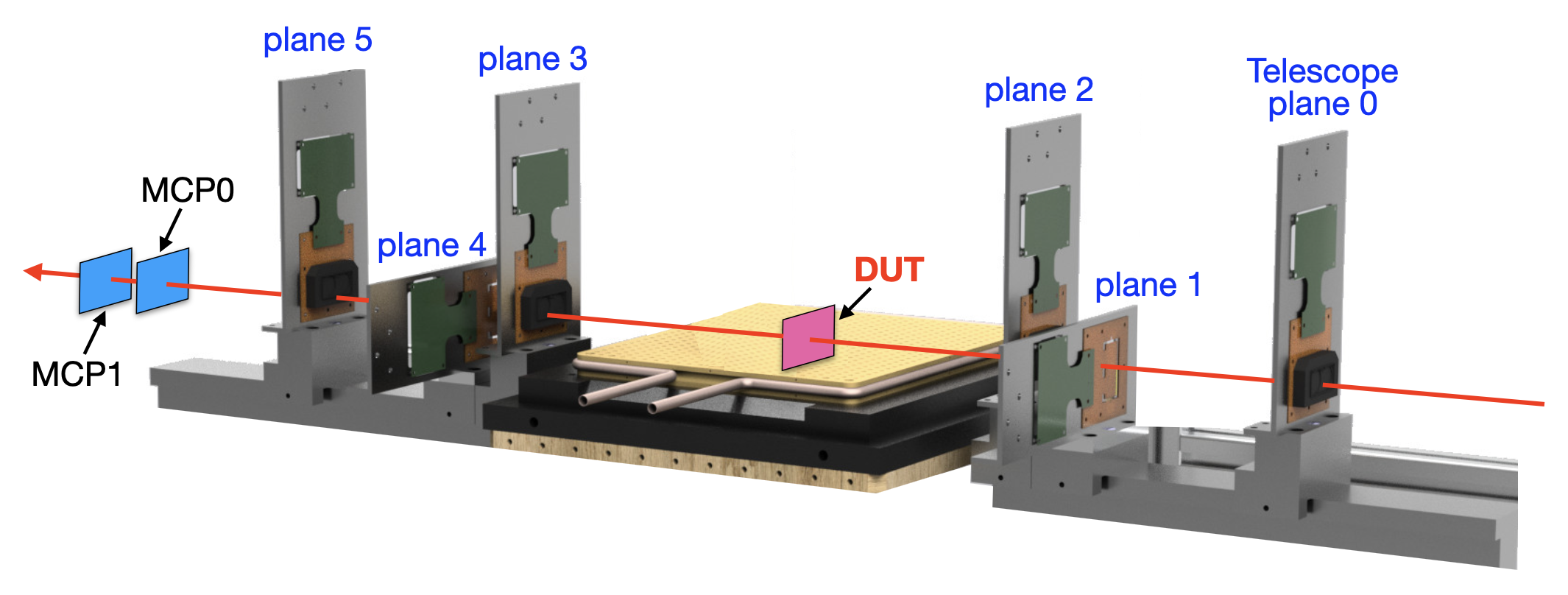}
\caption{\label{fig:Setup} Schematic view of the experimental setup, showing the six planes of the UNIGE FEI4 telescope \cite{FEI4_telescope}, the DUT, and the two MCPs installed downstream the telescope with respect to the pion beam direction.}
\end{figure}

The experimental setup is illustrated in Figure~\ref{fig:Setup}. The device under test (DUT), wire-bonded on a dedicated PCB board, was installed in the middle of the six detection planes of the UniGe FEI4 telescope for charged-particle tracking \cite{FEI4_telescope}. Two commercial MCP detectors, MCP0 and MCP1 hereafter, with expected time resolution for minimum-ionizing particles of approximately 5 ps, were installed downstream the telescope to provide a very precise time reference.

Analog signals from all detectors were read by two oscilloscopes. A first oscilloscope with a sampling rate of 40 GS/s and  an analog bandwidth set to 4 GHz was used to read the two MCP detectors and the differential output from pixel OA0 of the DUT. Pixels OA1, OA2 and OA3 of the DUT were instead connected to a second oscilloscope with 4 GHz analog bandwidth and a sampling rate of 20 GS/s. The oscilloscopes produced a saturation of the signal amplitudes to around 75 mV for all the DUT pixels. 

The FEI4 telescope   provided the trigger signal to the oscilloscopes, with a region of interest of $250 \times 250 $~\si{\um\squared} imposed to the first telescope plane and centered at the pixel OA0.
 A window of 200 ns   around the telescope trigger time was defined to record waveform data from the oscilloscopes.  

\subsection{Telescope-Track Selection Criteria}\label{selection}

Pion trajectories and their point of intersection with the DUT were reconstructed using the hits recorded by the FEI4 telescope planes. 
To maximise the telescope pointing resolution, tracks were required to have hits in all six telescope planes
and $ \chi^{2}/NDF \le 1.5 $.
To suppress ambiguous cases, events with more than one reconstructed track were rejected.
For tracks satisfying the aforementioned criteria, the telescope pointing resolution on the DUT plane was estimated in~\cite{mateus_thesis} to be approximately \SI{10}{\um}.

\subsection{Data samples}

Large samples of data were acquired at the four sensor bias voltage values $HV$ = 120, 160, 200 and 250 V, and at the five power density values reported in Table~\ref{tab:gain}.
For all these working points, the samplings of the differential signal in the first 30 ns of the trigger time window were used to determine the standard deviation of the voltage noise $\sigma_V$ of the DUT at the output of the analog front end. 

\begin{figure}[!htb]
\centering
\includegraphics[width=.49\textwidth]{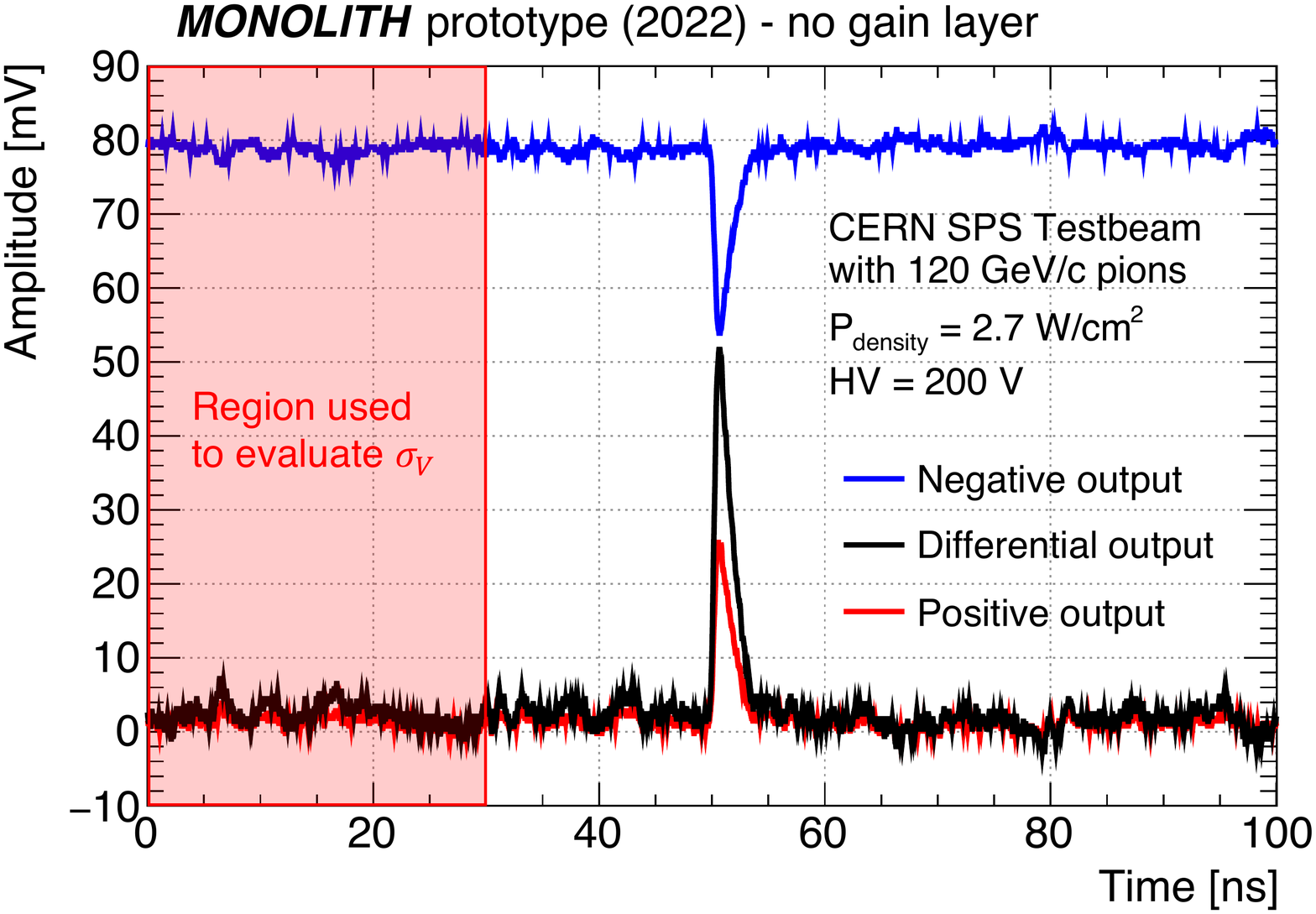}
\includegraphics[width=.49\textwidth]{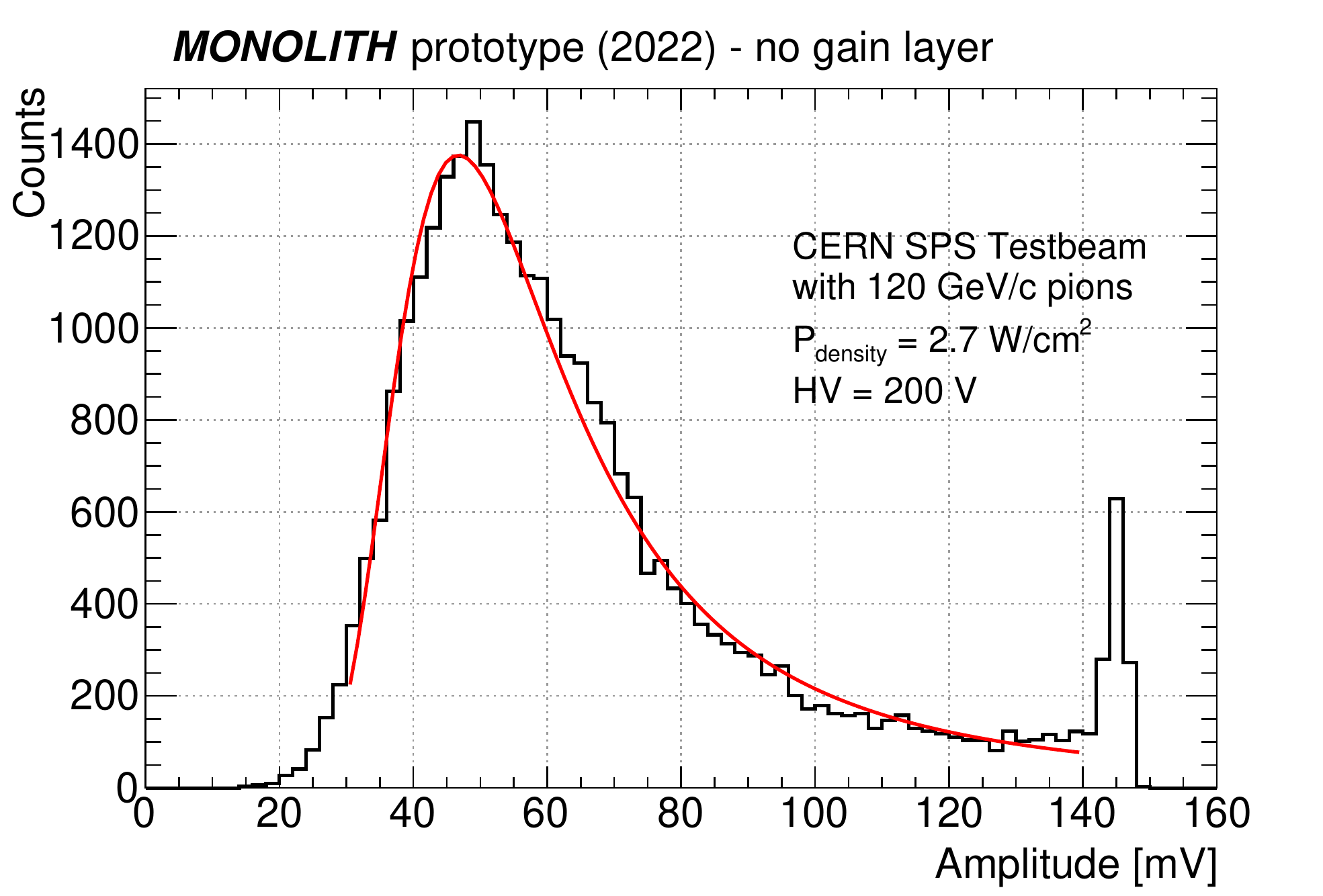}
\caption{\label{fig:waveform} Characteristics of signals from  pixel OA0 with the DUT operated at power density $P_{\mathrm {\it density}} =$ 2.7 W/cm$^2$ and bias voltage $HV=\SI{200}{\volt}$.
(Left) Example of waveforms from the single-ended outputs with the two polarities (blue and red) and from the differential output (black). 
The analog signal amplitudes from pixel OA0 were sampled every $\SI{25}{\pico\second}$. 
The voltage noise $\sigma_V$ was extracted taking the standard deviation of the differential signal samplings acquired in the time interval between 0 and 30 ns. 
(Right) Distribution of the amplitudes of the differential signals; the peak at large values is generated by the saturation of the oscilloscope. The red line is the result of a fit using a Landau functional form.}
\end{figure}

The left panel of Figure~\ref{fig:waveform}  shows a typical waveform acquired for pixel OA0 of the DUT operated at power density $P_{\mathrm {\it density}} =$ 2.7 W/cm$^2$ and bias voltage $HV=\SI{200}{\volt}$.
The distribution of the differential-signal amplitudes from pixel OA0 after the telescope-track selection is shown in the right panel of Figure~\ref{fig:waveform}.

\section{Detection Efficiency Measurement}
\label{sec:eff}

The pion tracks reconstructed by the FEI4 telescope and surviving the analysis selection criteria were used to measure the DUT detection efficiency. The efficiency was computed as the ratio between the number of selected tracks associated to recorded signals with amplitudes above a threshold of 7 times the voltage noise $\sigma_V$ and the total number of selected tracks traversing the DUT in the area corresponding to the four analog pixels. A tolerance of $\SI{10}{\um}$ in the region outside the external boundaries of the four-pixel area was allowed to account for the finite telescope pointing resolution.

\begin{figure}[!htb]
\centering
\includegraphics[width=.49\textwidth,trim=0 0 0 0]{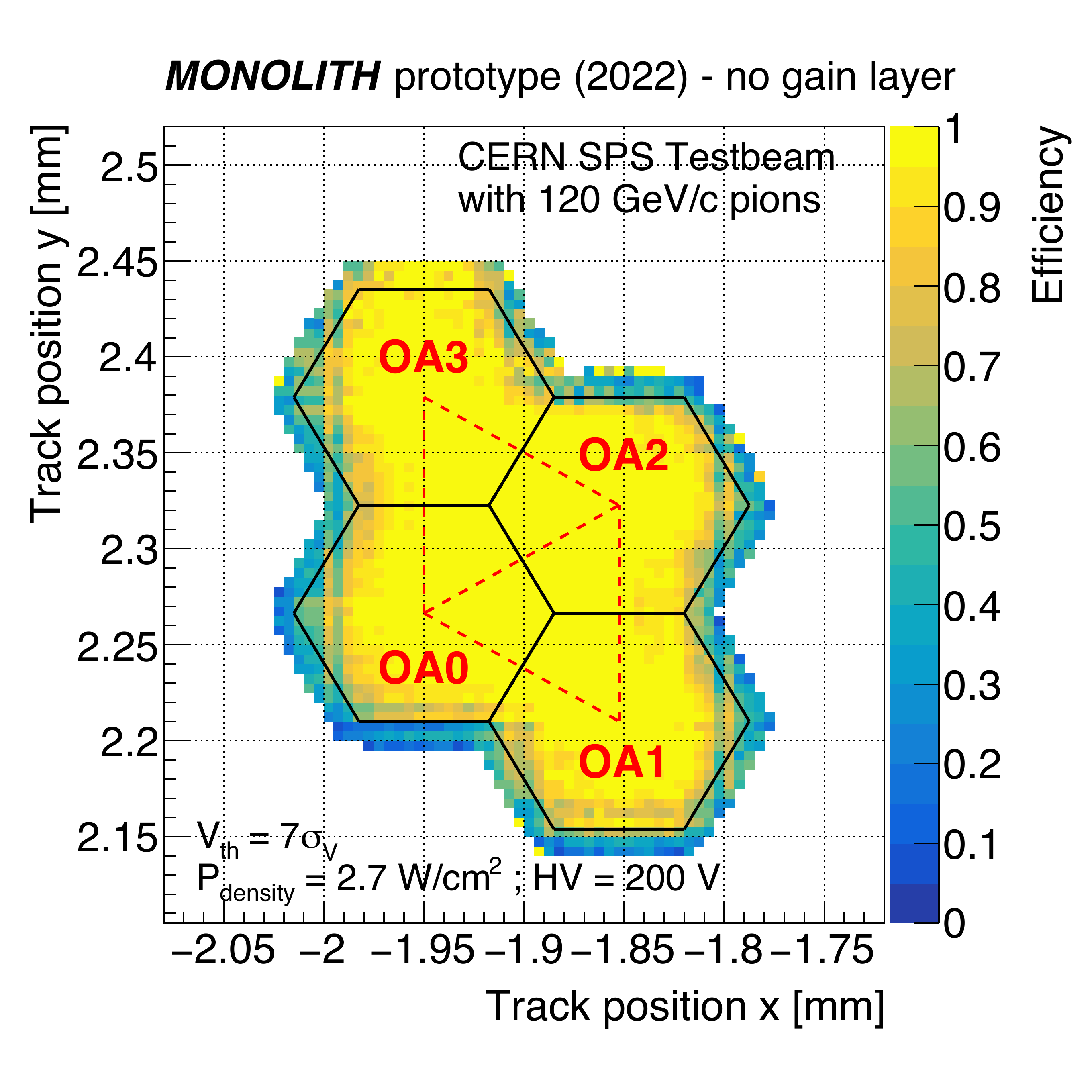}
\includegraphics[width=.49\textwidth,trim=0 0 0 0]{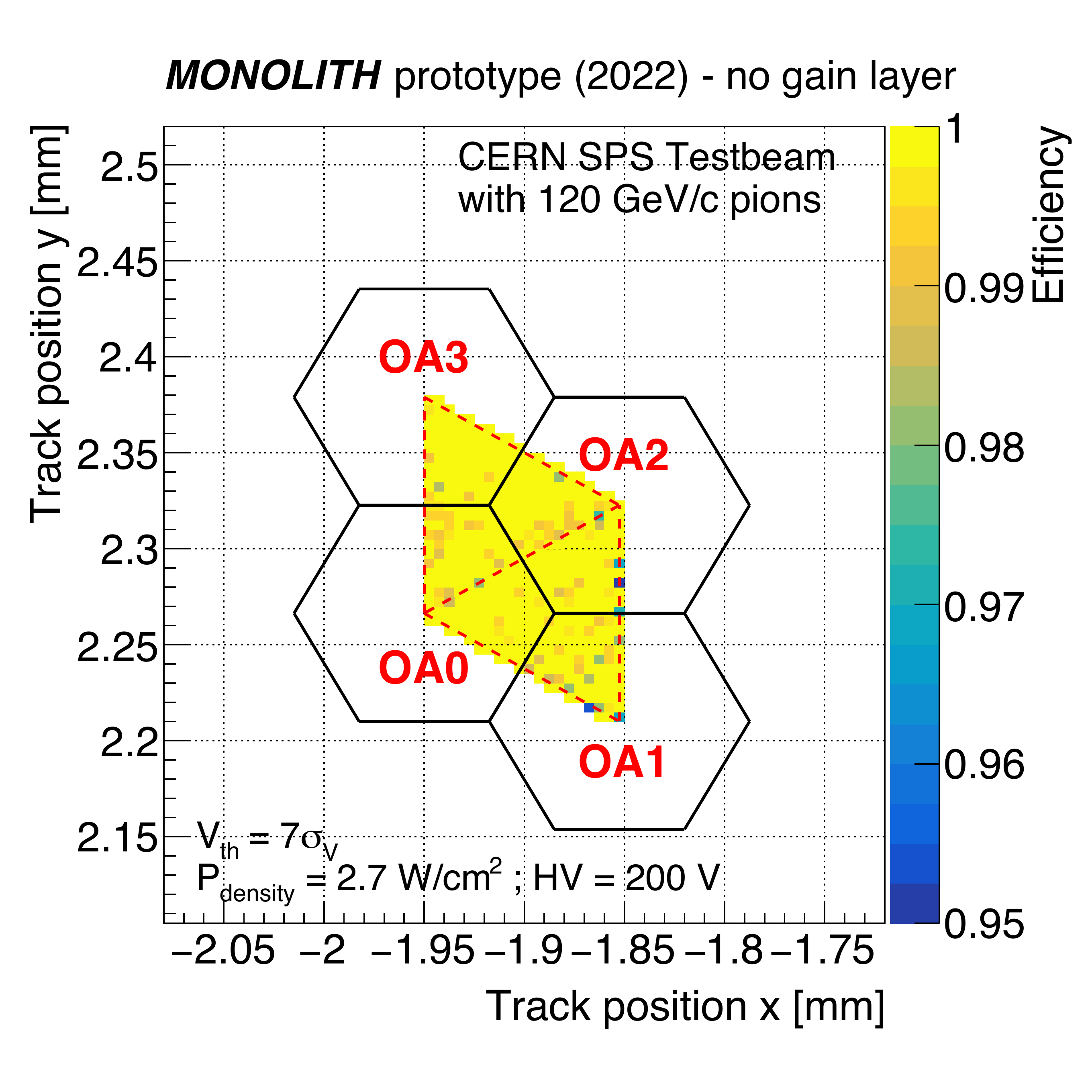}
\caption{\label{fig:effmap} Detection efficiency measured for the \monolith\ prototype operated at power density $P_{\mathrm {\it density}} =$ 2.7 W/cm$^2$ and $HV=\SI{200}{\volt}$, with a discrimination threshold $V_{\mathrm {\it th}}=7~\!\sigma_V$. In the left panel, the measurement is performed in a region that extends beyond the pixel edges (black lines) by $\SI{10}{\um}$. The apparent decrease of the efficiency  at the external boundaries delimiting the area of the four instrumented pixels is produced by the finite FEI4 telescope pointing resolution. 
The right panel shows the efficiency measured in the two triangular areas connecting the centres of pixels OA0--OA2--OA3 and OA0--OA2--OA1, which is unaffected by the resolution of the telescope; the colour scale of the plot starts at 95\% efficiency. 
}
\end{figure}

Figure~\ref{fig:effmap} shows the results of the efficiency measurement for the DUT operated at $P_{\mathrm {\it density}} =$ 2.7 W/cm$^2$ and $HV=\SI{200}{\volt}$. As shown by the left panel, the finite FEI4 telescope pointing resolution  
generates an apparent inefficiency close to the 14 external edges of the four pixels (and an efficiency outside them), which makes the evaluation of the efficiency in the total pixel area difficult.
Since this effect is not present in the five internal edges of the four pixels,
the measurement was repeated selecting only tracks extrapolated inside the two triangles connecting the OA0--OA2--OA3 and OA0--OA2--OA1 pixel centres. 
The result is shown in the right panel of Figure~\ref{fig:effmap}, in which the color scale starts at 95\%.
It should be noted that the two triangles contain all the relevant pixel regions, such as the boundary between two pixels and the corner between three pixels, in the same proportion as in an entire pixel. For all the working points, the efficiencies measured separately in the two triangles were found to be compatible within statistics, and thus averaged.
The efficiency value measured inside the two triangles for the working point of Figure~\ref{fig:effmap} is ($ 99.86_{-0.04}^{~\!+0.03} $)\%, as reported in the second row of Table~\ref{tab:eff_ipream_HV_pscan}.

\begin{table}[!htb]
\centering 
\renewcommand{\arraystretch}{1.4}
\begin{tabular}{|c|c|c|}
\cline{1-3}
 $P_{\it density}$ [W/cm$^2$] & ~~~~~~~$HV$~[\si{V}]~~~~~~~  & Average efficiency [\%] \\
\cline{1-3}
2.7   & 250 & $ 99.80_{-0.09}^{~\!+0.07} $ \\
2.7   & 200 & $ 99.86_{-0.04}^{~\!+0.03} $ \\
2.7   & 160 & $ 99.71_{-0.10}^{~\!+0.08} $ \\
2.7   & 120 & $ 99.80_{-0.09}^{~\!+0.07} $ \\
0.9   & 200 & $ 99.83_{-0.04}^{~\!+0.04} $ \\
0.36  & 200 & $ 99.81_{-0.05}^{~\!+0.04} $ \\
0.13  & 200 & $ 99.77_{-0.05}^{~\!+0.04} $ \\
0.04  & 200 & $ 99.86_{-0.07}^{~\!+0.05} $ \\
\cline{1-3}
\end{tabular}
\caption{Detection efficiency measured for the \monolith\ 2022 prototype at different power density   and sensor bias voltage values. A voltage threshold of 7 times the voltage noise $\sigma_V$ is applied. The efficiencies are measured separately in the two triangular areas connecting the OA0--OA2--OA3 and OA0--OA2--OA1 pixel centers shown in Figure~\ref{fig:effmap}, and then averaged.}
\label{tab:eff_ipream_HV_pscan}
\end{table} 

\begin{figure}[!h]
\centering 
\includegraphics[width=.70\textwidth,trim=0 0 0 0]{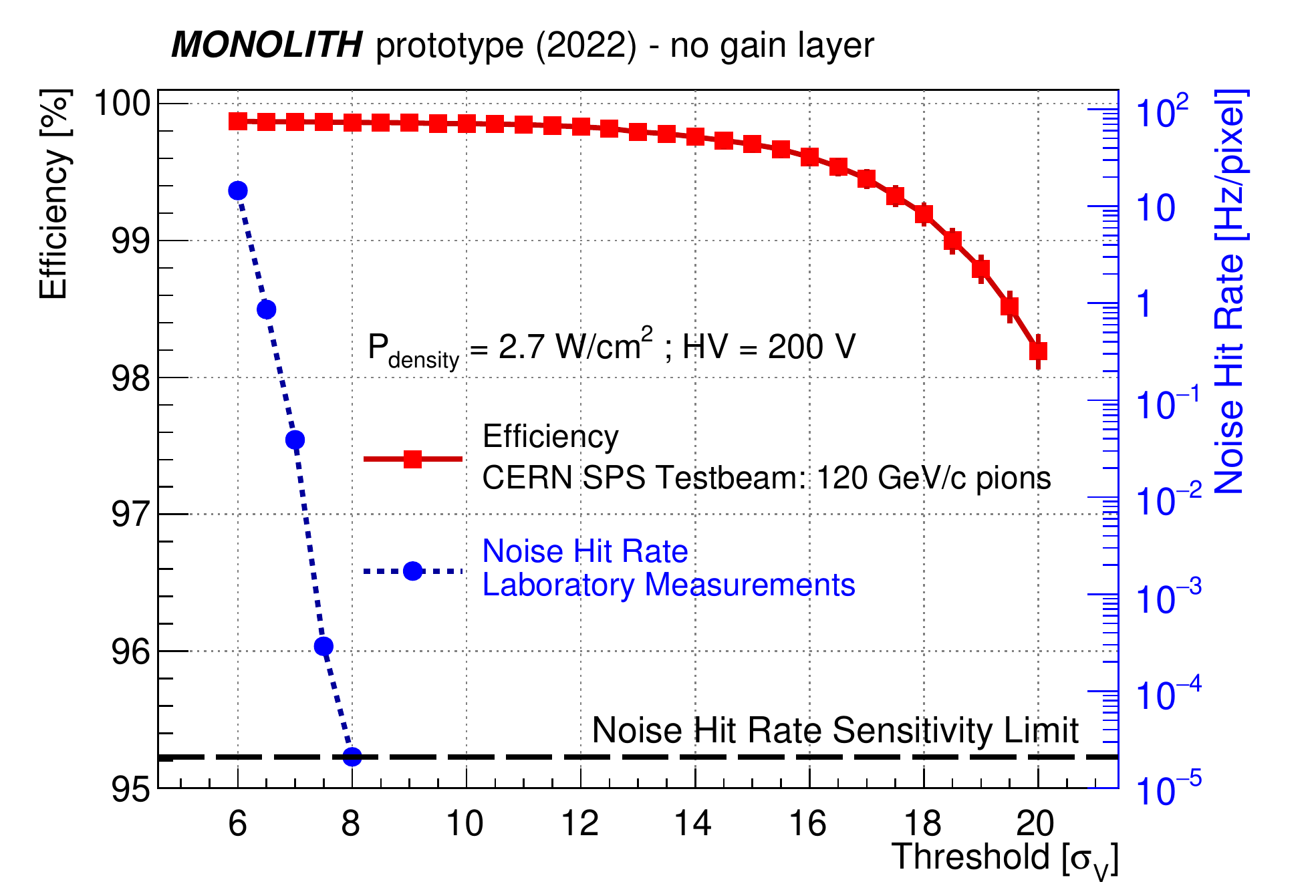}
\caption{\label{fig:effthrscan} Detection efficiency (red squares) and noise-hit rate (blue squares) measured within the two triangular areas depicted in Figure~\ref{fig:effmap}, shown as a function of the voltage threshold in integer multiples of the the voltage noise $\sigma_V$. The data refer to sensor bias $ HV = \SI{200}{\volt} $ and power density $P_{\mathrm {\it density}} =$ 2.7 W/cm$^2$.
}
\end{figure}

It is interesting to characterise the dependence of the detection efficiency on the discrimination threshold, as it is influenced by the signal-to-noise ratio of the  sensor prototype. The result for the DUT operated at $P_{\mathrm {\it density}} =$ 2.7 W/cm$^2$ and $HV=\SI{200}{\volt}$ is shown in Figure~\ref{fig:effthrscan}. Remarkably, the efficiency remains above $99\%$ up to threshold values of 18 times the voltage noise $\sigma_V$. 
A voltage threshold of $7~\!\sigma_V$ was used throughout the data analysis, which corresponds to approximately 600 electrons for this working point.
Figure~\ref{fig:effthrscan} reports also the noise-hit rate measured in the laboratory, which is $4\cdot 10^{-2}$ Hz/pixel at $7~\!\sigma_V$.
In the testbeam data sample taken at this working point, we expect less than 1 noise hit.

\begin{figure}[!htb]
\centering 
\includegraphics[width=.49\textwidth,trim=0 0 0 0]{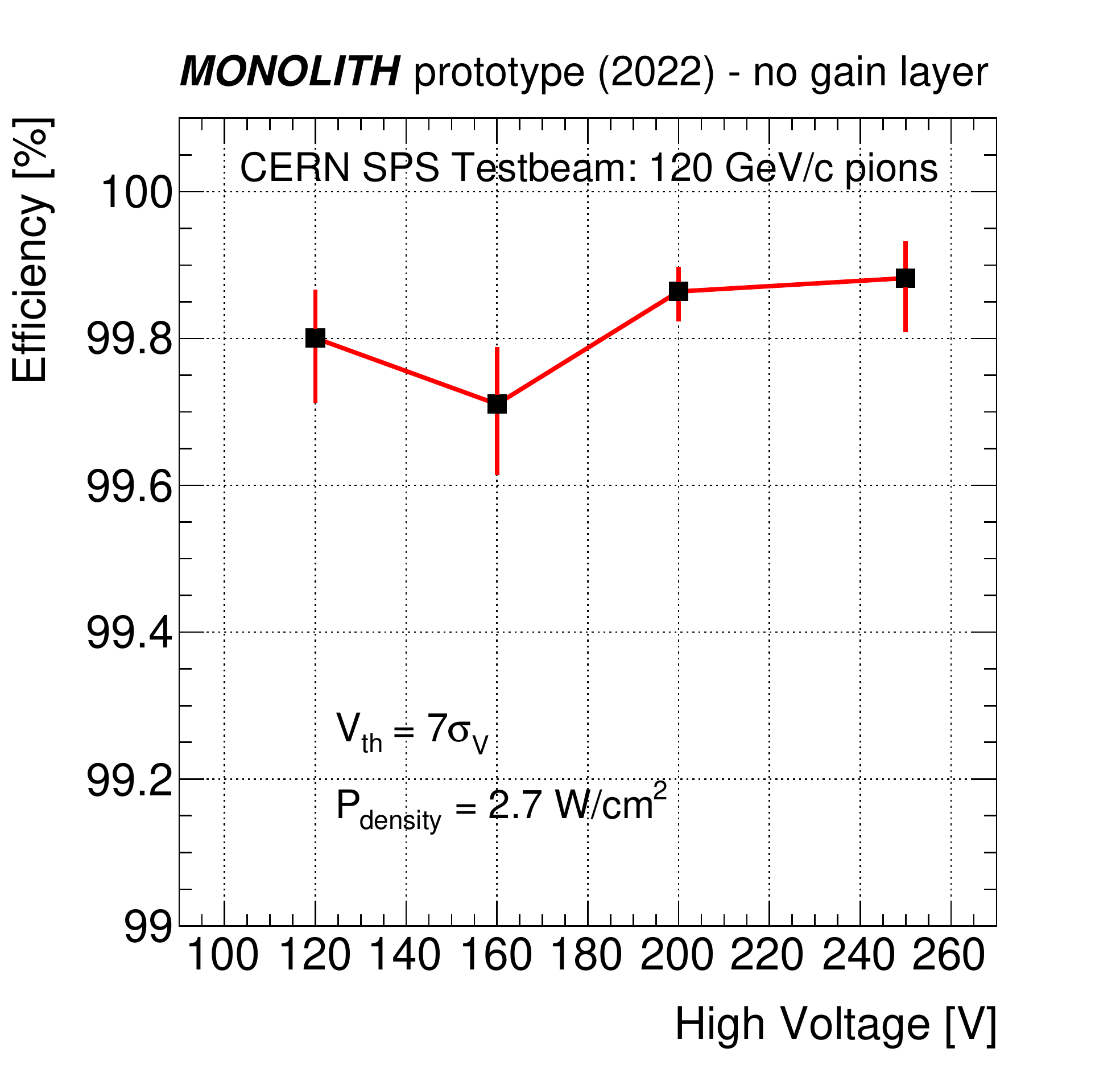}
\includegraphics[width=.49\textwidth,trim=0 0 0 0]{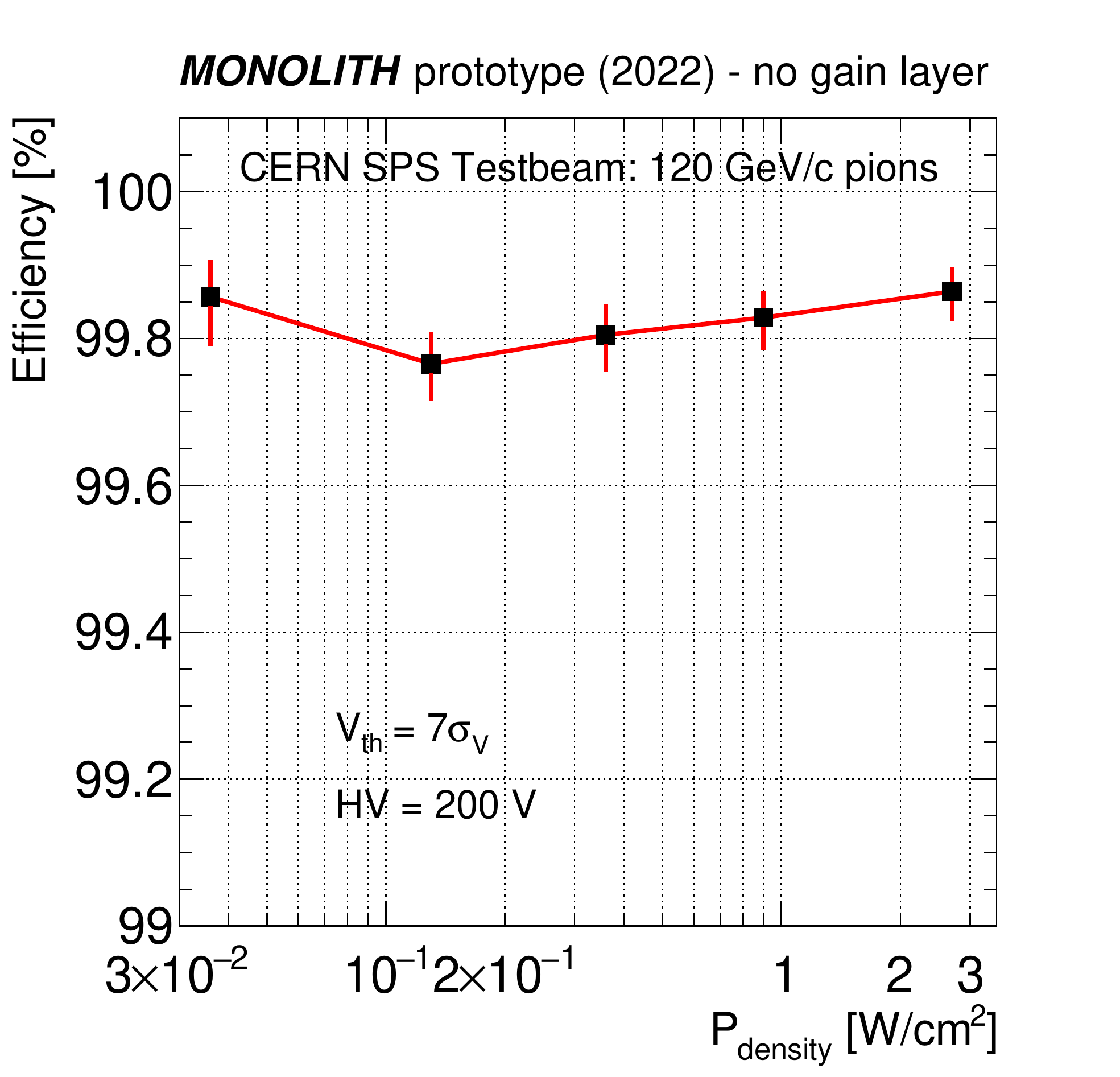}
\caption{\label{fig:eff_ipream_HV_pscan} Detection efficiency measured within the two triangular areas depicted in Figure~\ref{fig:effmap} with $V_{\mathrm {\it th}}=7~\!\sigma_V$. In the left panel, the efficiency is measured at four values of sensor bias voltage for a power density of 2.7 W/cm$^2$. In the right panel, the measurement is shown for five power density values at sensor bias $HV=\SI{200}{\volt}$.}
\end{figure}

\begin{figure}[!htb]
\centering 
\includegraphics[width=.49\textwidth,trim=0 0 0 0]{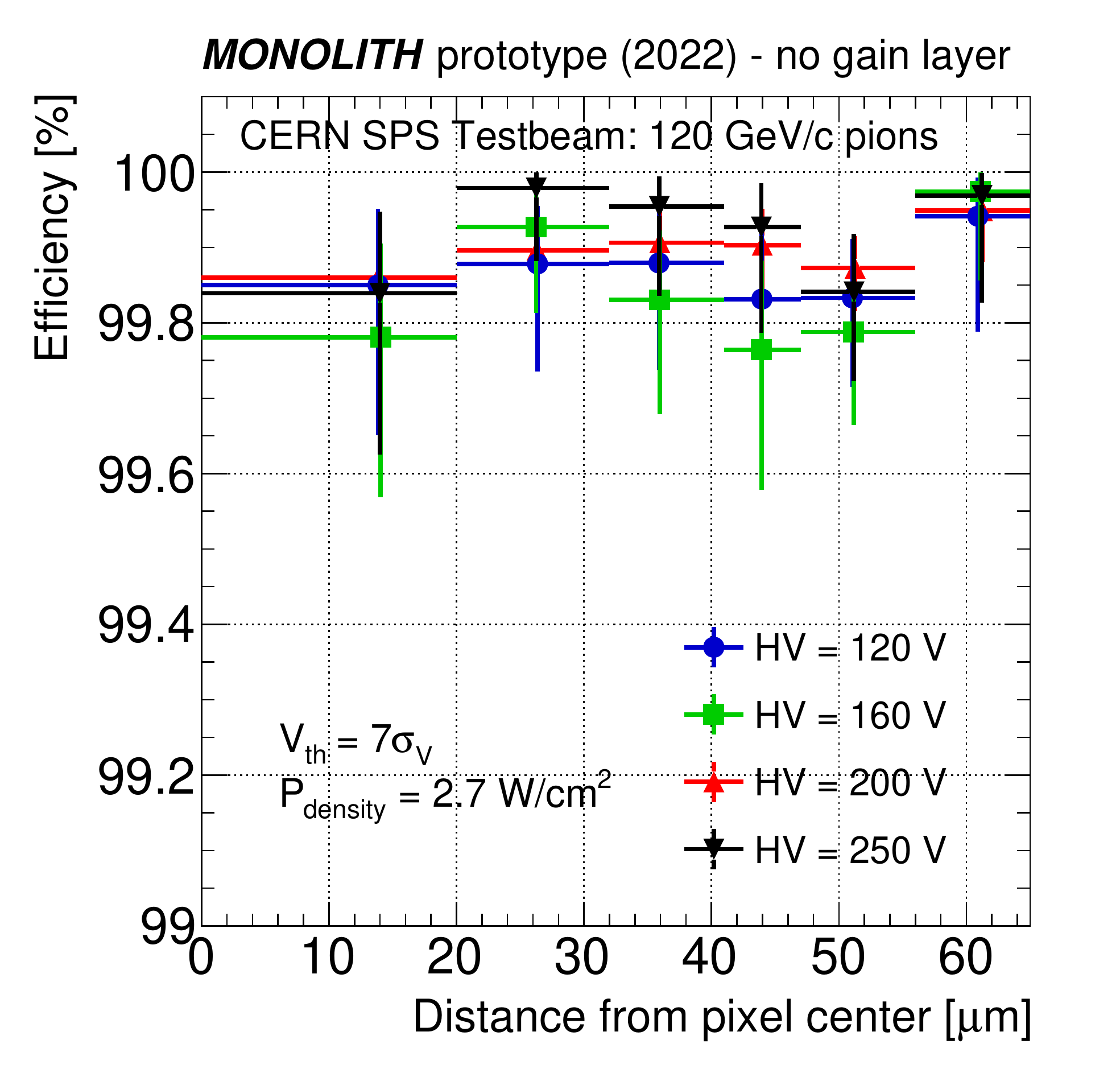}
\includegraphics[width=.49\textwidth,trim=0 0 0 0]{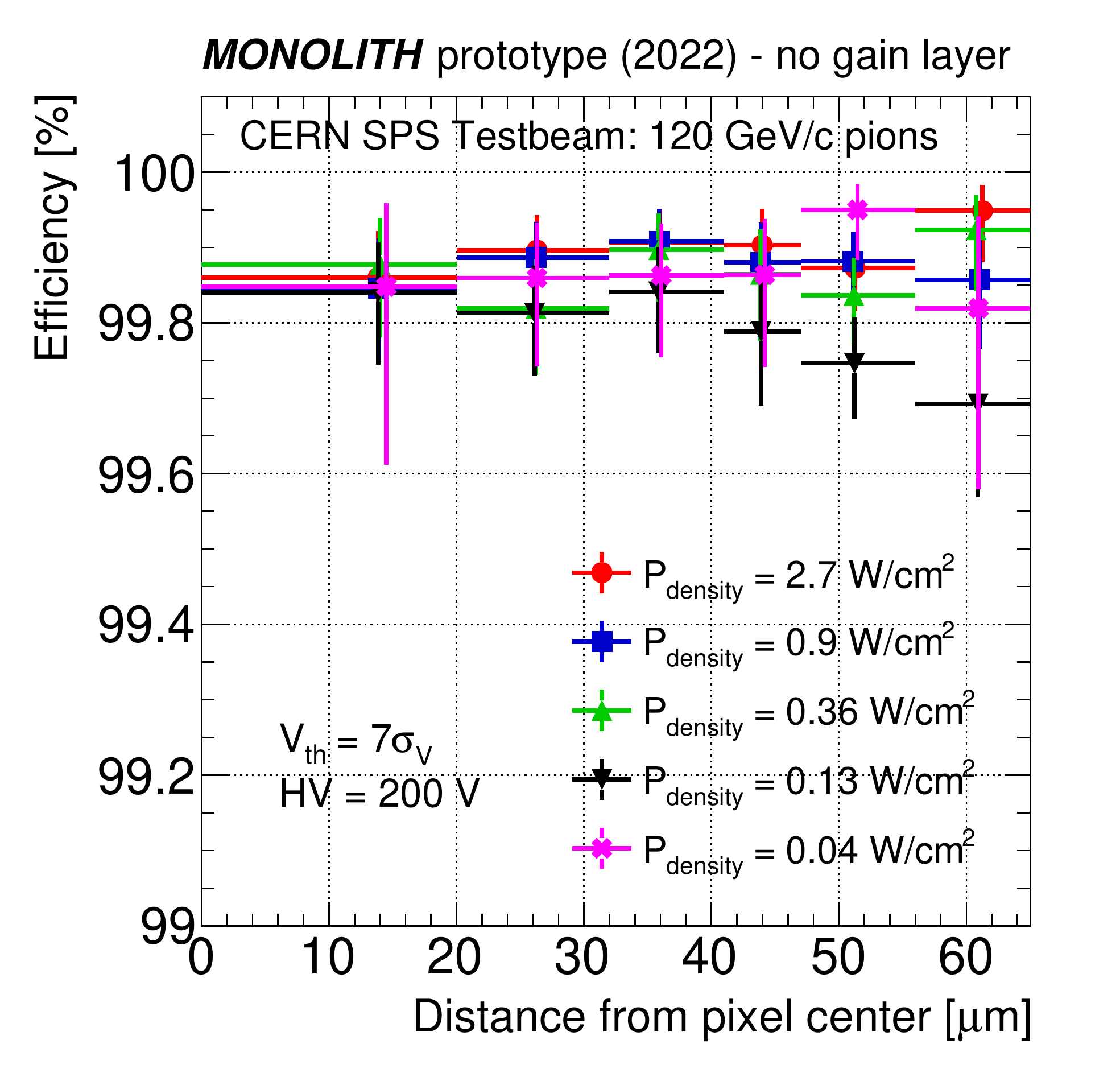}
\caption{\label{fig:eff_radius} Detection efficiency measured within the two triangular areas depicted in Figure~\ref{fig:effmap}, in six distance intervals from the  pixel center. In the left panel, the  efficiency is shown for four values of $HV$ at a power density of 2.7 W/cm$^2$. In the right panel, it is shown for five power density values at sensor bias $HV=\SI{200}{\volt}$.  In each bin the data-point marker is placed at the average distance value from the pixel center.
}
\end{figure}

Table~\ref{tab:eff_ipream_HV_pscan} and Figure~\ref{fig:eff_ipream_HV_pscan} show the average efficiency in the two triangles for the eight working points at which data were acquired. 
Within the statistical uncertainty, no significant trend is observed: all the efficiencies are found to be compatible with 99.8\%. Figure~\ref{fig:eff_radius} shows  the DUT efficiencies measured as a function of the distance from the pixel center for the eight working points acquired. In all cases the efficiencies remain remarkably stable around 99.8\% within the whole pixel area, showing that no drop in the inter-pixel region is measured.

\section{Time Resolution Measurement}
\label{sec:reso}
The time measurements of the DUT and the two MCPs for the sample of telescope tracks selected as described in Section~\ref{selection} were used to determine the timing performance of the \monolith\ 2022 prototype. 
The measurement was restricted to events with tracks crossing the area of pixel OA0 inside the two triangles shown in Figure~\ref{fig:effmap}, to exclude hits in the inter-pixel region that have the majority of the deposited charge  in the three non-instrumented pixels adjacent to OA0, and thus should not be associated to pixel OA0. 

The DUT time resolution was studied using the time of arrival ($\toa$) of the  analog signals acquired by pixel OA0, which was connected to the fastest oscilloscope. 
The $\toa$ values for the DUT were computed as the time at which the linear interpolation between two consecutive oscilloscope samplings of the differential signal reached the 7$~\!\sigma_V$ threshold. 
The difference in the measured $\toa$ ($\dtoa$) between the three possible detector pairs, DUT-MCP0, DUT-MCP1 and MCP0-MCP1, corresponds to the time-of-flight between the detectors. Under the assumption of a Gaussian  response of the three detectors, the standard deviations extracted from the corresponding $\dtoa$ distributions, $\sigdtoa$, are interpreted as the sum in quadrature of the time resolution of the two detectors in the pair. By solving a system of three equations, one for each $\sigdtoa$ value, the three unknown time resolutions $\sigdut$, $\sigmcpzero$ and $\sigmcpone$ can therefore be retrieved.

\subsection{Time-walk correction}
\begin{figure}[!h]
\centering
\includegraphics[width=.49\textwidth,trim=0 0 0 0]{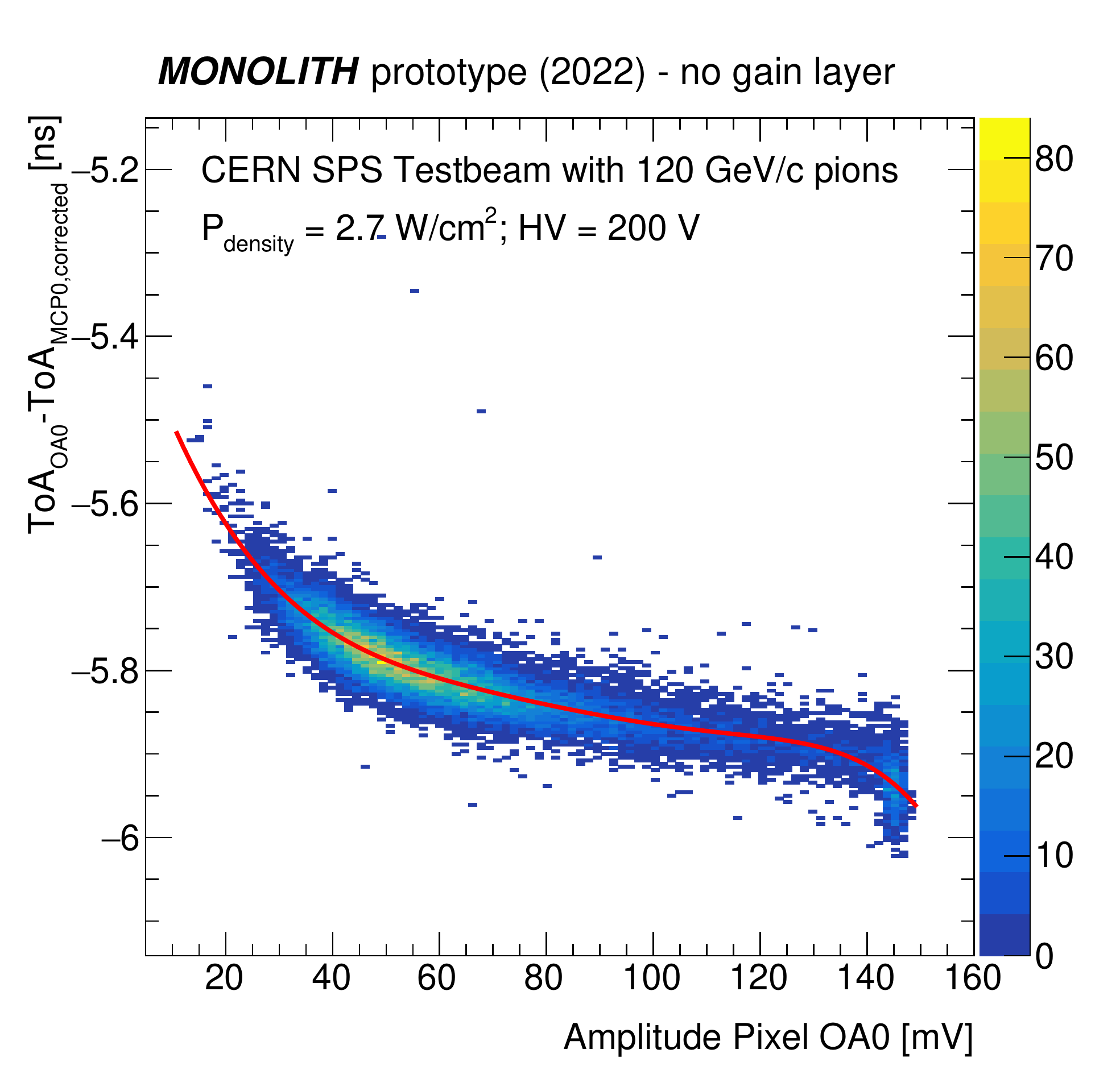}
\includegraphics[width=.49\textwidth,trim=0 0 0 0]{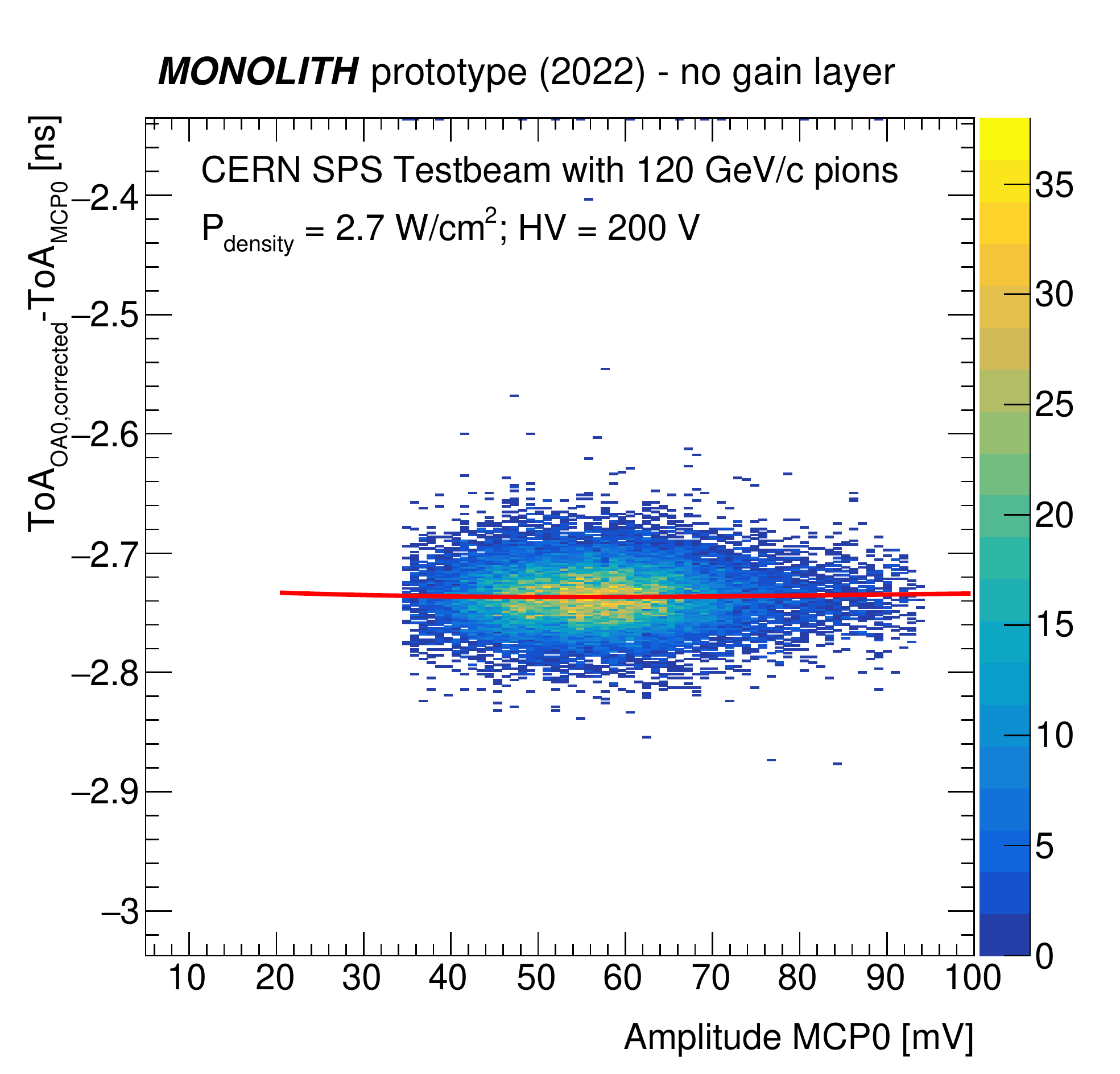}
\caption{\label{fig:twcorr} Distribution of the difference $\dtoa$ between the DUT and MCP0 as a function of the DUT signal amplitude (left) and the MCP0 signal amplitude (right).
The DUT was operated at power density of 2.7 W/cm$^2$ and sensor bias voltage of $\SI{200}{\volt} $. An arbitrary offset with no influence on the measured time resolution is present in the $\dtoa$. 
The red lines show the time-walk corrections obtained with the unbinned maximum likelihood fit method described in the text. The corrections were parametrised with polynomial functions with the coefficients determined simultaneously for the three detectors.}
\end{figure}

The distributions of the $\dtoa$ as a function of the amplitudes of thedetectors in the pair was used to correct for time walk.
As an example, Figure~\ref{fig:twcorr} shows the $\dtoa$ distribution for the DUT and MCP0 pair as a function of the DUT (left) and MCP0 (right) signal amplitudes
for the working point with power density of 2.7 W/cm$^2$ and $HV = \SI{200}{\volt}$. 
Two independent methods were deployed and cross-checked against each other to exclude potential biases to the measured timing performance introduced by the time-walk correction procedure. 

\begin{itemize}
\item The first correction method is equivalent to the technique used to obtain the results documented in~\cite{PicoAD_TB}. A Gaussian fit to the $\dtoa$ distribution was performed to determine its most probable value, separately for events falling in bins of the signal amplitude. 
The values obtained were 
linearly interpolated, resulting in a continuous parametrisation of the correction as a function of the signal amplitude. Each event was then corrected two times, once for each of the detectors involved in the $\dtoa$. 

\item The second method relies on an unbinned maximum likelihood fit to extract the time-walk corrections, modeled with polynomial functions $f_\text{DUT}$, $f_\text{MCP0}$ and $f_\text{MCP1}$ of the corresponding signal amplitudes, simultaneously for all detectors. A likelihood estimator $\mathcal{L}$ was constructed under the assumption of a Gaussian time response from all sensors.  The probability of observing a $\dtoaid$ difference for a selected track $i$ and the detector pair $d$ can thus be described with a Gaussian probability density function $\mathcal{G}_{i,d}$. In the model, the standard deviation parameter $\sigdtoad$ describes the sum in quadrature of the time resolutions of the two detectors in pair $d$. Additionally, being $d1$ and $d2$ the two detectors involved in the $d$ pair associated to $\dtoaid$, the total time-walk correction for track $i$ is given by $\mu_{i,d}\equiv f_{d2}(\mathrm{Amplitude}_{d2,i})-f_{d1}(\mathrm{Amplitude}_{d1,i})$. All the parameters of the model were determined by maximizing
\begin{equation}
\mathcal{L}=\log\left(\prod_i^{\text{tracks}} \prod_d^{\text{detector}\atop\text{pairs}} \mathcal{G}_{i,d}( \dtoaid|\sigdtoad,\mu_{i,d})\right).
\end{equation}
A sixth-order polynomial function was chosen for the DUT, in order to model the effect of the amplitude saturation, while third-order polynomials parametrised the time-walk effect for the MCPs. The advantage of this method is the absence of an arbitrary choice of the binning, whilst it is limited by the capability of modeling precisely the functional dependence of the time-walk correction on the signal amplitude.
\end{itemize}

These two independent methods yielded results in excellent agreement, with observed differences well below 1 ps, confirming that systematic effects stemming from the extraction of the time-walk correction are marginal.

\subsection{Results}
\label{sec:twc}

For the calculation of the time resolutions, Gaussian fits of the  time-walk corrected $\dtoa$ distributions of the three pairs of detectors were performed independently for each working point.
To avoid possible biases of the time-resolution values coming from the non-Gaussian tails, the range of the fit was limited to the bulk of each $\dtoa$ distribution by including only bins populated by more than 25\% of the entries in the bin with the largest content. The $\sigma$ parameters of the Gaussian fits were then used to solve the system of equations that yields the time resolutions for the DUT OA0 and the two MCPs. 

\begin{figure}[!htb]
\centering %
\includegraphics[width=.33\textwidth,trim=40 20 30 20]{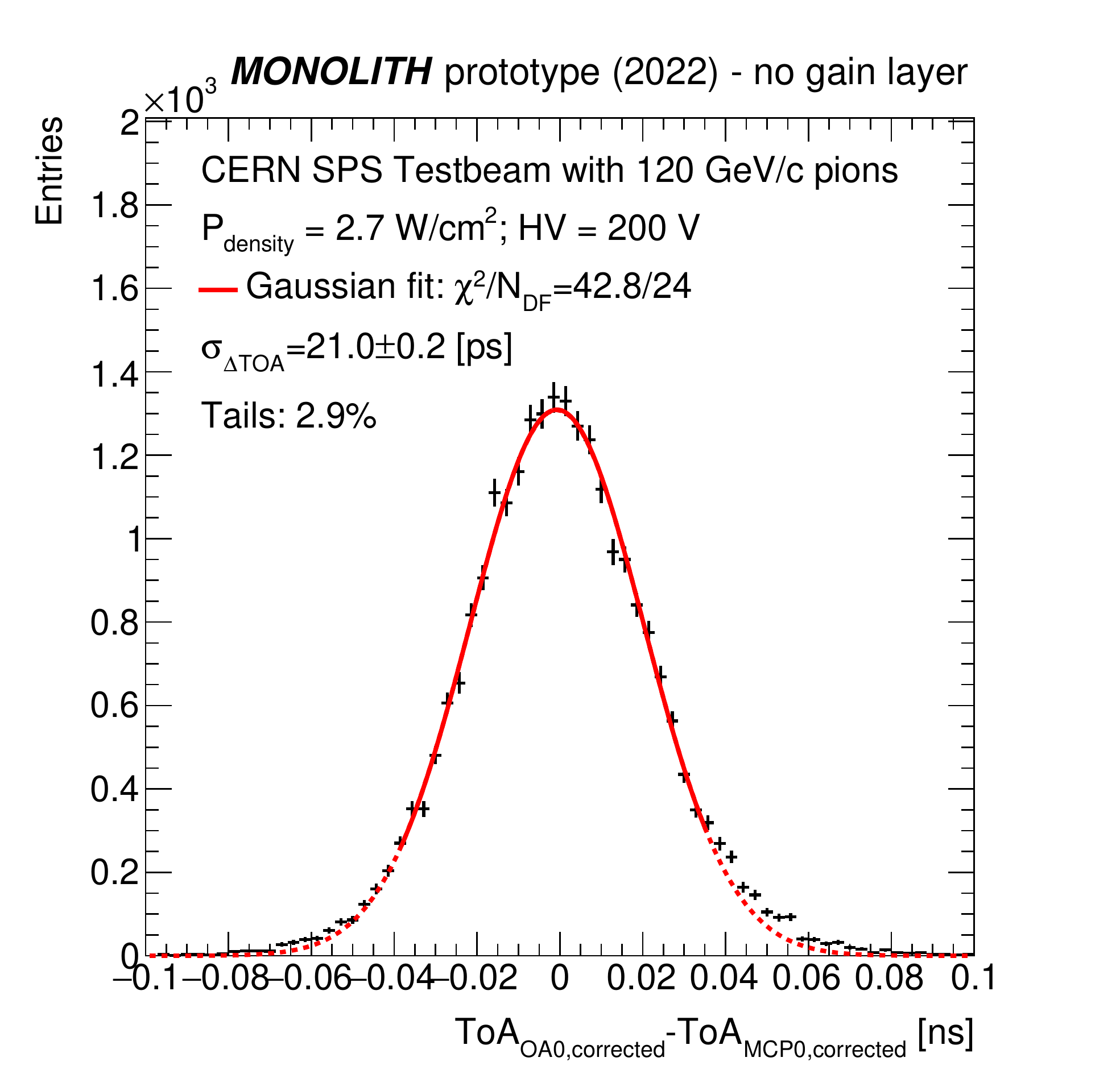}
\includegraphics[width=.33\textwidth,trim=40 20 30 20]{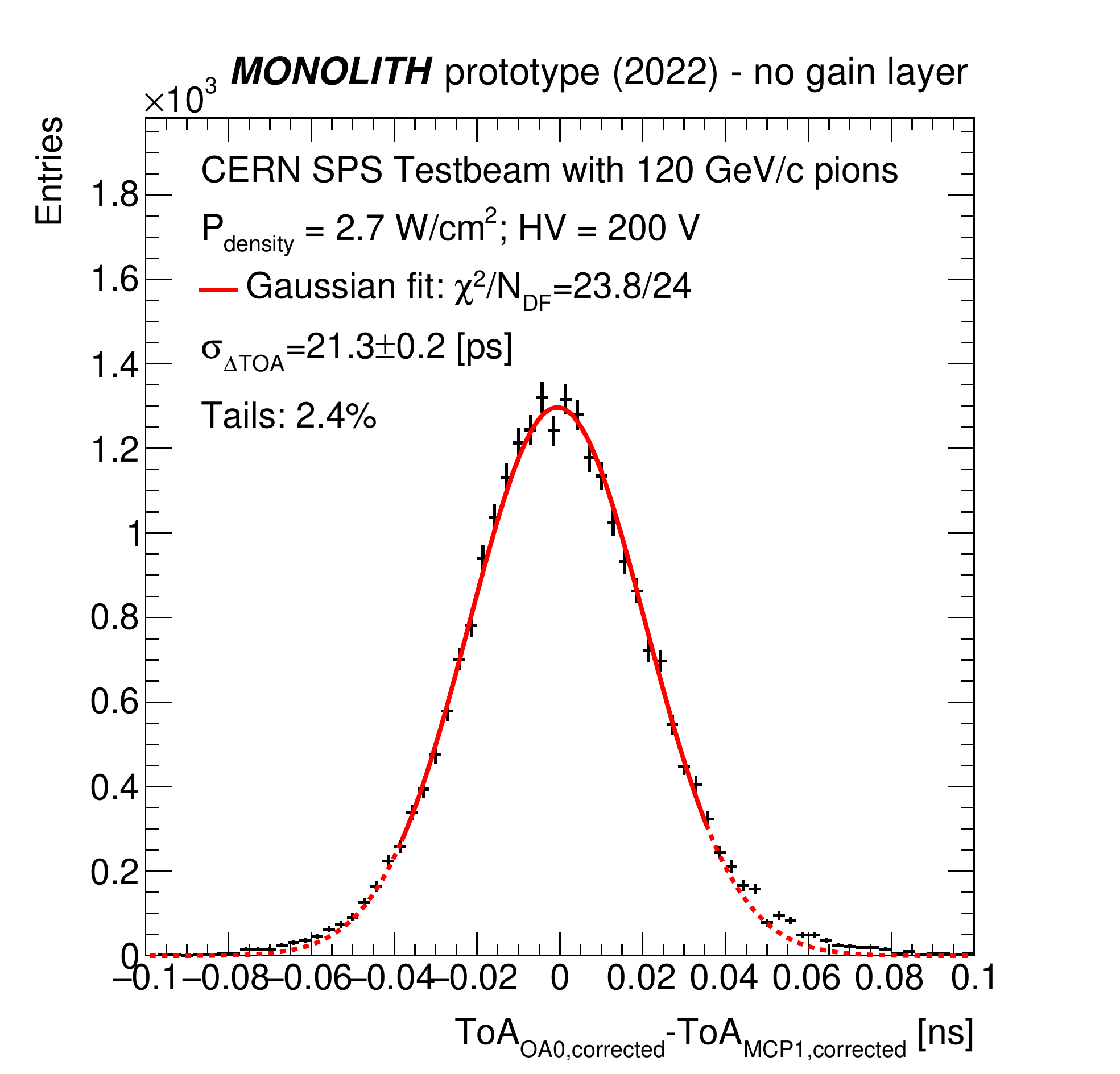}
\includegraphics[width=.33\textwidth,trim=40 20 30 20]{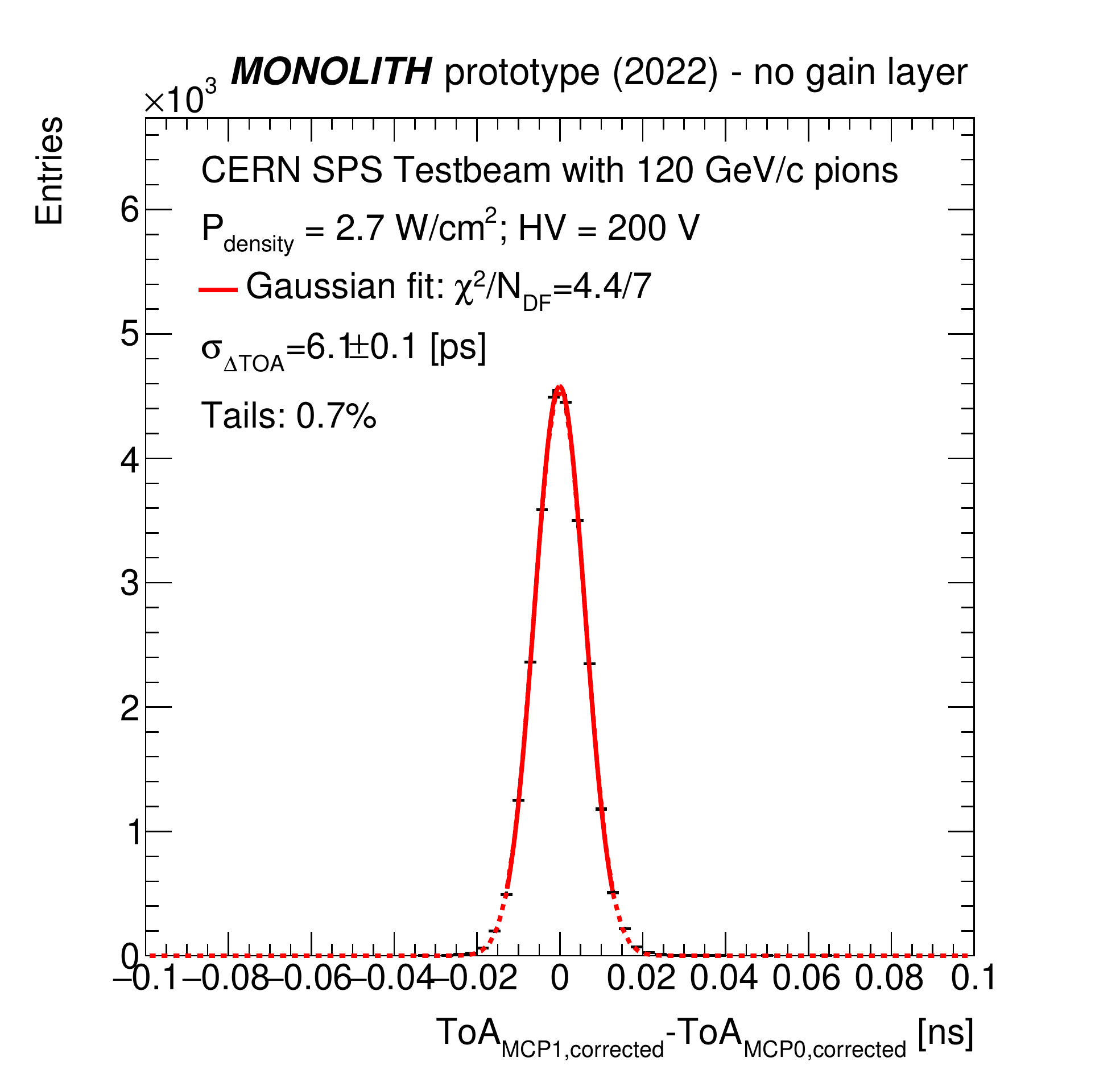}
\caption{\label{fig:TOF_fits} Distributions of the time-walk-corrected $\dtoa$ difference between the DUT and MCP0 (left), DUT and MCP1 (center), and between MCP0 and MCP1 (right), for the working point with power density of 2.7 W/cm$^2$ and $ HV = \SI{200}{\volt} $.
The full red lines show the results of the Gaussian fits to the bulk of each distribution, while the dashed lines extrapolate the fit to the entire histogram range. The non-Gaussian contributions in the tails are reported in the plots, as well as the three $\sigdtoa$ used to calculate the DUT and MCPs time resolutions.}
\end{figure}

As an example,  the  $\dtoa$ distributions for the working point with power density of 2.7 W/cm$^2$ and sensor bias voltage $ HV = \SI{200}{\volt} $ are shown in Figure~\ref{fig:TOF_fits}.
A time resolution of $(20.7\pm 0.3)$ ps is measured 
for this working point.
For the same working point, a time resolution of 48 ps was measured without the correction for time walk.

The very narrow distribution of the $\dtoa$ between the two MCPs in Figure~\ref{fig:TOF_fits} indicates their excellent timing performance, measured to be ($3.6\pm1.5$) ps for MCP0 and ($5.0\pm1.1$) ps for MCP1, and confirms their ability to serve as a good time reference. 

The total fraction of events exceeding the Gaussian-fit integral in the $\dtoa$ distributions involving the DUT was found to be below 3\% for all the working points, showing that the non-Gaussian component of the time response of the \monolith\ 2020 prototype is overall small and that the  resolutions measured with this method describe at least 97\% of the recorded signals. 
It was verified that these small non-Gaussian tails  mainly originate from events with small amplitudes that are associated to pions crossing the DUT in the inter-pixel regions, for which  pixel OA0 collected only a fraction of the charge produced.
Extension of the time-resolution measurement to all events containing a hit in pixel OA0 (thus removing the request for the hit to be within the two triangles of Figure~\ref{fig:effmap}) does not change the time resolution values. The only effect produced is that the events with $\dtoa$ outside the Gaussian fits increases from 3\% to 5\%; it was checked that these extra events in the tails are associated to hits in the inter-pixel region of the three sides of the OA0 pixel hexagon for which the adjacent pixels were not readout.

\begin{figure}[!htb]
\centering %
\includegraphics[width=.49\textwidth,trim=0 0 0 0, clip]{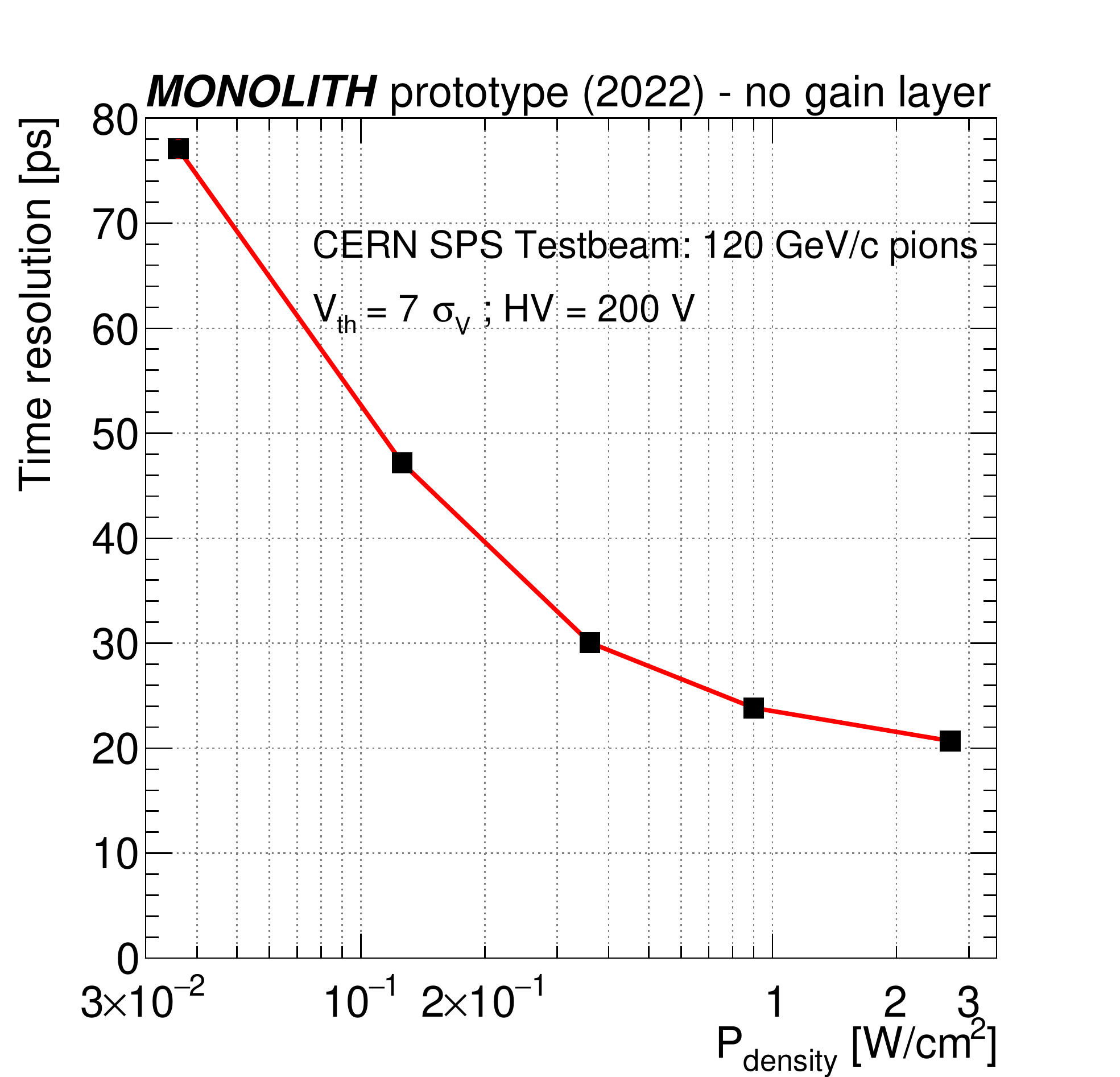}
\includegraphics[width=.49\textwidth,trim=0 0 0 0, clip]{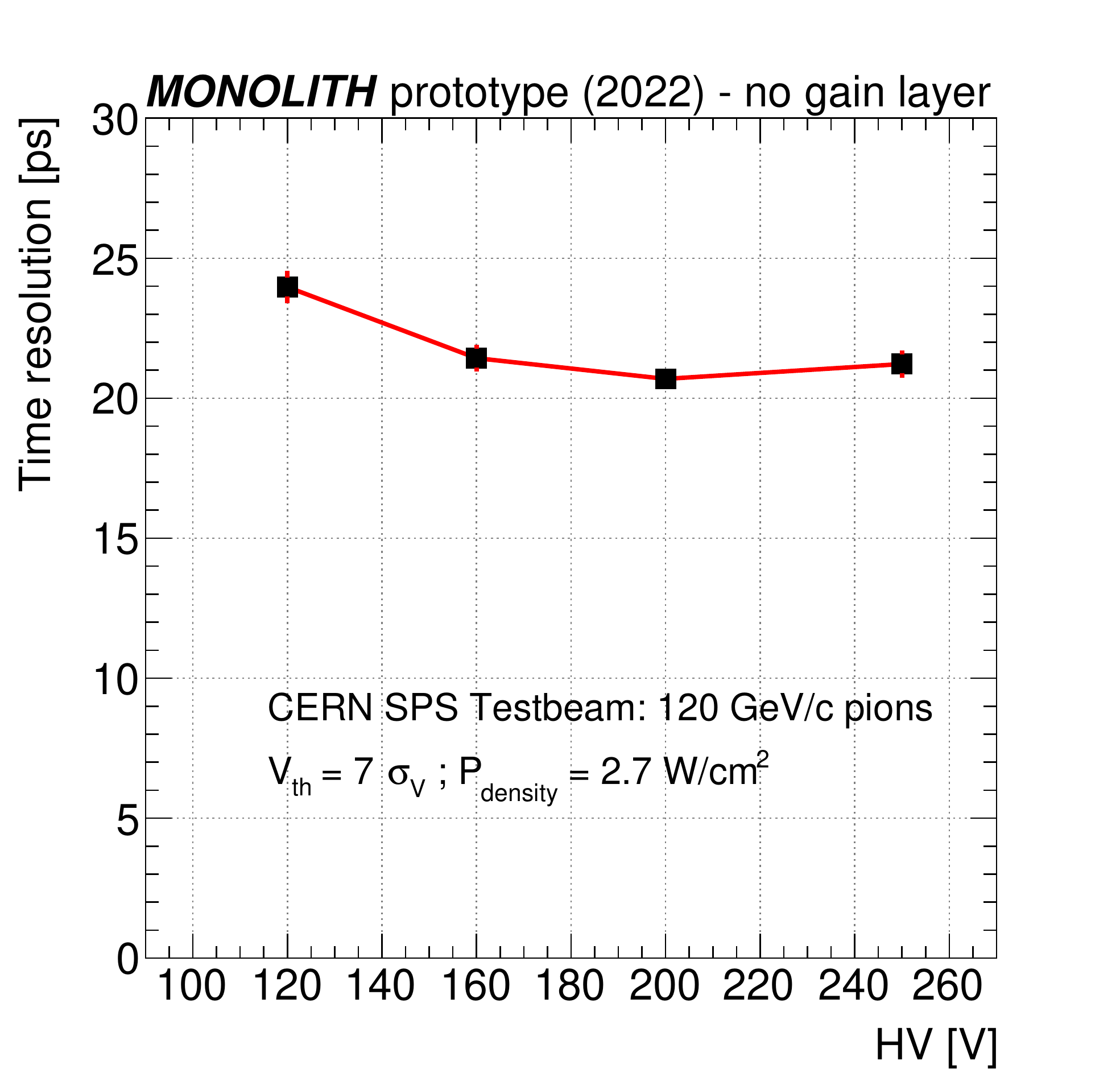}
\caption{\label{fig:TOF_power_HV} Time resolution measured for sensor bias voltage HV = 200 V as a function of the power density  (left panel), and for power density of 2.7 W/cm$^2$ as a function of sensor bias voltage (right panel). }
\end{figure}

Table~\ref{tab:tabsumm} and the left panel of Figure~\ref{fig:TOF_power_HV} report the time resolution measured at  $HV$ = 200 V
for the five power density values ranging from 0.04 to 2.7 W/cm$^2$. 
While the efficiency remains at the 99.8\% level in all cases, a gradual deterioration of the timing performance is visible at lower power density values, although it remarkably remains at the level of $ \SI{30}{\pico\second}$ at  power density of 0.36 W/cm$^2$ and $ \SI{80}{\pico\second}$ at 0.04 W/cm$^2$. 

\begin{table}[!htb]
\centering
\renewcommand{\arraystretch}{1.2}
\begin{tabular}{|c|c|c|}
\cline{1-3}
\multicolumn{3}{|c|}{DUT operated at  $ HV = \SI{200}{\volt} $ and $V_{\it th} = 7 \sigma_V$} \\ 
\cline{1-3}
 $~~P_{\it density}$ [W/cm$^2$]~~ & Amplitude MPV [mV] & Time Resolution [ps] \\
\cline{1-3}
2.7   & $ 48.6 \pm 0.5 $ &   $20.7 \pm 0.3$ \\
0.9   & $ 35.8 \pm 0.5 $ &   $23.8 \pm 0.3$ \\
0.36  & $ 22.6 \pm 0.4 $ &   $30.1 \pm 0.4$ \\
0.13  & $ 14.2 \pm 0.3 $ &   $47.2 \pm 0.7$ \\
0.04  & $ 16.2 \pm 0.3 $ &   $77.1 \pm 0.9$ \\

\cline{1-3}
\end{tabular}
\caption{
Time resolution at sensor bias voltage $HV = \SI{200}{\volt}$ for the five power consumption per unit surface values studied. 
The measurements refer to the area of pixel OA0 that is inside the two triangles of Figure~\ref{fig:effmap}. 
The most probable values of the amplitude of the differential signals are also reported.
The uncertainties are statistical only.
}
\label{tab:tabsumm} 
\end{table}

The right panel of Figure~\ref{fig:TOF_power_HV} shows the time resolution
 as a function of the sensor bias voltage for power density $P_{\mathrm {\it density}} =$ 2.7 W/cm$^2$. The measurement show that the DUT can be operated in a wide $HV$ range with a time resolution between 20 and 25 ps.

\begin{figure}[!htb]
\centering %
\includegraphics[width=.49\textwidth,trim=0 0 0 0, clip]{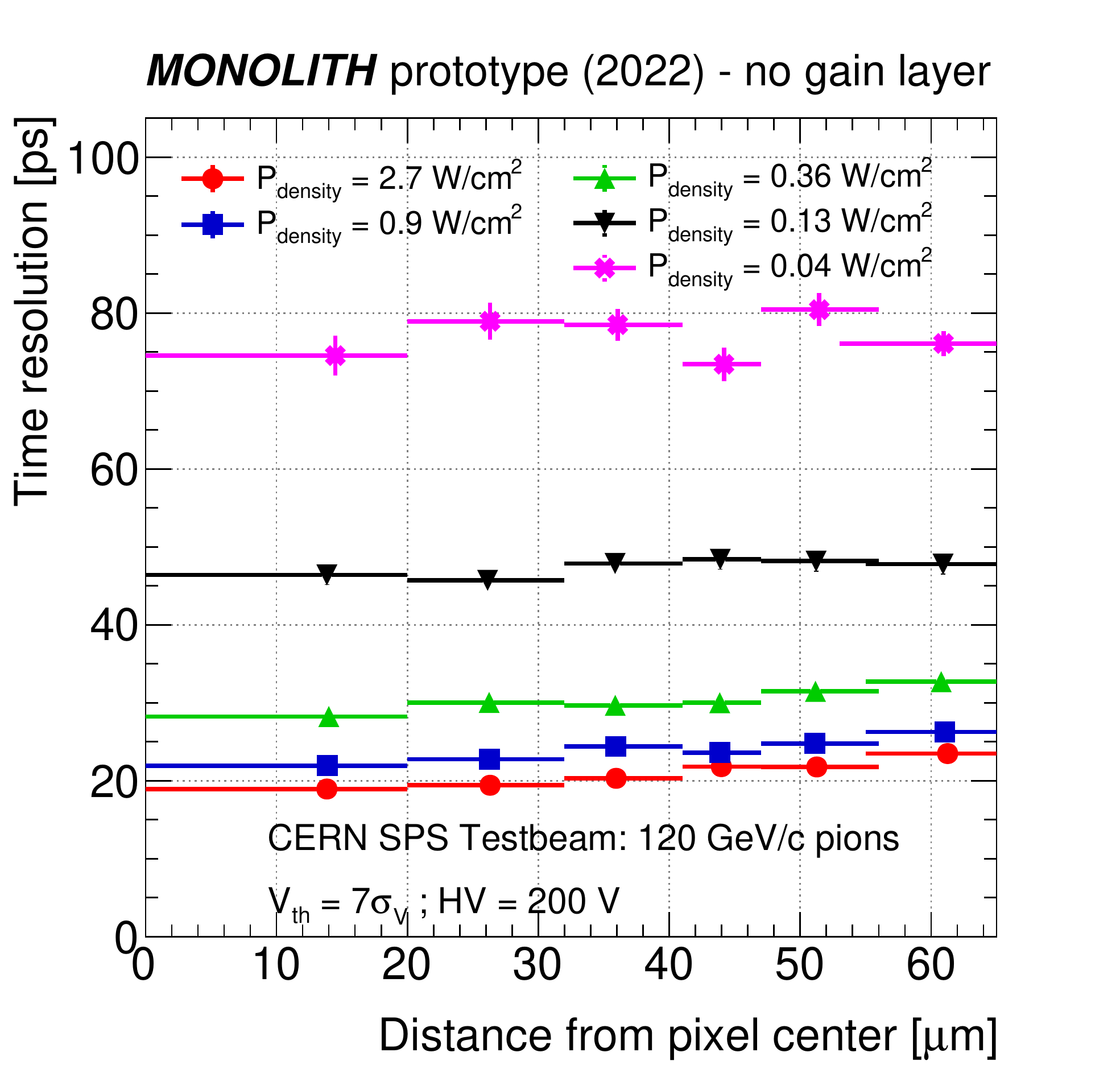}
\includegraphics[width=.49\textwidth,trim=0 0 0 0, clip]
{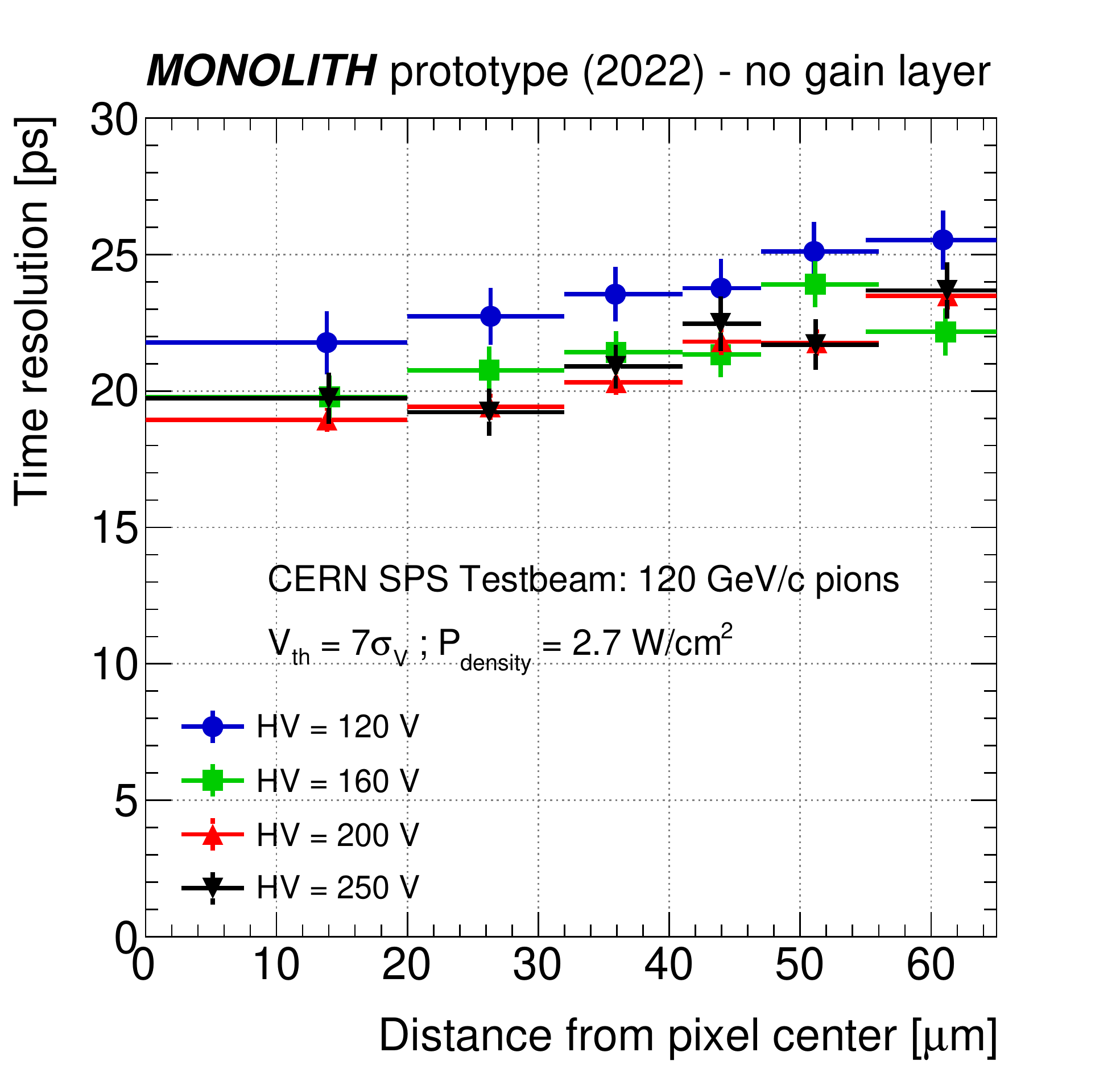}
\caption{\label{fig:TOF_radius} 
Time resolution as a function of the distance from center of the pixel OA0. In the left panel, the  resolution is shown for the five power density values acquired at sensor bias voltage HV = 200 V. In the right panel, it is measured at a power density of 2.7 W/cm$^2$ for the four sensor bias voltage values acquired. In each  bin the data point marker is placed at the average distance value from the pixel center.}
\end{figure}

The time resolution was also measured as a function of the distance from the center of the pixel OA0. 
%
The left panel of Figure~\ref{fig:TOF_radius} shows that a mild dependence on the distance from the pixel center is measured only for $\pd \ge$ 0.36 W/cm$^2$, when the time resolution is below $\approx$30 ps.
The right panel of the Figure shows that the time resolutions measured for $HV$ between 160 and 250 V are compatible with each other within uncertainties, while at $HV$ = 120 V the measurements are systematically higher by approximately 2 ps. This observation might hint that the charge drift velocity at $HV$ = 120 V is not yet saturated deep in the  sensor volume.
%

All the time resolution measurements are obtained with the $\toa$ computed with a simple threshold setting (that offers the significant advantage of requiring a simple electronics circuitry) and a simple signal-processing method (linear interpolation between consecutive oscilloscope samplings).
More complex signal-processing methods (mimicking a low-pass RC filter, a constant-fraction discriminator, or spline interpolation of the oscilloscope signal samplings), which would require  more complex electronics, improve marginally the time resolutions, at most at the 15\% level.

\section{Conclusions}
\label{sec:conclusions}

The monolithic silicon pixel matrix prototype produced in 2022 with the SG13G2 SiGe BiCMOS IHP process in the framework of the H2020 MONOLITH ERC Advanced project was tested with a beam of 120 GeV/c pions at the CERN SPS. The sensor in this ASIC has a depletion depth of 50 µm and does not incorporate an internal gain layer.

Samples of events at several sensor bias voltages and front-end power consumption were acquired.
The detection efficiency using a voltage threshold seven times the voltage noise is measured to be consistent with 99.8\% for all the working points analyzed. This efficiency does not depend on the position of the hit within the pixel surface, demonstrating very large efficiency also in the inter-pixel regions.

The differential signals from analog channels were sampled and acquired during the testbeam using fast oscilloscopes. 
The time resolution obtained computing the signal time of arrival at the voltage threshold by a simple linear interpolation of the acquired signal samplings,  which would require a rather simple electronics circuitry, 
is measured to be 20 ps at $HV$ = 200 V and front-end power consumption of 2.7 W/cm$^2$. 
More elaborated signal-processing procedures, corresponding to more complex and performing electronics, improve the time resolution by at most 15\%.
The prototype can be operated with a time resolution of 30 ps at 0.36 W/cm$^2$ and 80 ps at 0.04 W/cm$^2$. 
The measured time resolutions have little dependency on the sensor bias voltage  in the range from 120 to 250 V.
They also have little dependency on the position of the hit within the pixel area, 
with  a small degradation towards the edge of the pixel.

It is worth stressing that these remarkable results are obtained with a monolithic silicon sensor without internal gain layer.
\acknowledgments
This research is supported by the H2020 MONOLITH project, ERC Advanced Grant ID: 884447. The authors wish to thank Coralie Husi, Javier Mesa, Gabriel Pelleriti and all the technical staff of the University of Geneva and IHP microelectronics.
The authors acknowledge the support of EUROPRACTICE in providing design tools and MPW fabrication services.
\newpage
\bibliographystyle{unsrt}
\bibliography{bibliography.bib}
\end{document}